\documentclass[floatfix,twocolumn,preprintnumbers,amsmath,amssymb,superscriptaddress]{revtex4-2}
\usepackage[utf8]{inputenc}

\usepackage{amsmath}
\usepackage{amsfonts}
\usepackage{graphicx}
\usepackage[percent]{overpic} 
\usepackage{color}

\begin{document}

\title{Number of local minima in discrete-time fractional Brownian motion}
\author{Maxim Dolgushev}
\email{maxim.dolgushev@sorbonne-universite.fr}
\address{Laboratoire de Physique Th\'eorique de la Mati\`ere Condens\'ee,
CNRS/Sorbonne University, 4 Place Jussieu, 75005 Paris, France}
\author{Olivier B\'enichou}
\address{Laboratoire de Physique Th\'eorique de la Mati\`ere Condens\'ee,
CNRS/Sorbonne University, 4 Place Jussieu, 75005 Paris, France}
\date{\today}

\begin{abstract}
The analysis of local minima in time series data and random landscapes is essential across numerous scientific disciplines, offering critical insights into system dynamics. Recently, Kundu, Majumdar, and Schehr derived the exact distribution of the number of local minima for a broad class of Markovian symmetric walks [Phys. Rev. E \textbf{110}, 024137 (2024)]; however, many real-world systems are non-Markovian, typically due to interactions with possibly hidden degrees of freedom. This work investigates the statistical properties of local minima in discrete-time samples of fractional Brownian motion (fBm), a non-Markovian Gaussian process with stationary increments, widely used to model complex, anomalous diffusion phenomena. We derive a complete asymptotic characterization of the fluctuations of the number of local minima $m_N$ in an $N$-step discrete-time fBm. We show that the fluctuations of $m_N$ exhibit a sharp transition at the Hurst exponent $H=3/4$: for $H\le 3/4$ they satisfy a central limit theorem with Gaussian limiting law, whereas for $H>3/4$ they converge to a non-Gaussian Rosenblatt process. The convergence at the process level gives us full statistical description at all times. We exemplify it on the covariance of the rescaled minima process, which displays two qualitatively distinct regimes matching Brownian and Rosenblatt covariances on either side of this threshold. Our analysis relies on a Hermite/Wick decomposition of the local-minimum indicator, which isolates a quadratic functional of an effective long-memory mode as the unique driver of the anomalous statistics. These results identify the count of local minima as a simple and robust diagnostic of long-range dependence in non-Markovian Gaussian processes, a conclusion supported by numerical simulations.
\end{abstract}

\maketitle

The identification and analysis of local minima in time series and random landscapes play a fundamental role across various scientific fields \cite{masuda2025energy,Delage23}. Their occurrence and distribution often reveal crucial insights about the system, whether biological, environmental, financial, or physical. In biomedical applications, local minima in electrocardiogram signals are used to detect cardiac disorders \cite{sun2005characteristic} and to identify sleep phases in neuroimaging \cite{hasson2018combinatorial}.  Similarly, climate scientists aim to predict minimum daily temperatures for frost forecasting \cite{bhakare2025spatial}, financial analysts examine local extrema to determine market trends \cite{aichinger2010fast,chen2018estimation}, and in signal processing, local minima are used to decompose complex signals \cite{huang1998empirical}; finally, the analysis of local minima finds a wide range of applications in physics \cite{broderix2000energy,dayal2004performance,fyodorov2004complexity,bray2007statistics,ben2021counting,lacroix2022counting,gershenzon2023site,weinrib1982distribution,hivert2007distribution,bray1980metastable,annibale2003supersymmetric,lacroix2024superposition} --- for instance, the nummber of local minima in energy landscapes are critical to understanding phase transitions \cite{fyodorov2004complexity,bray2007statistics,ben2021counting,broderix2000energy,lacroix2022counting,gershenzon2023site,dayal2004performance}, as their abundance of such minima significantly influences the low-temperature behavior of spin glasses \cite{bray1980metastable,annibale2003supersymmetric}.

\begin{figure}
    \includegraphics[scale=0.29]{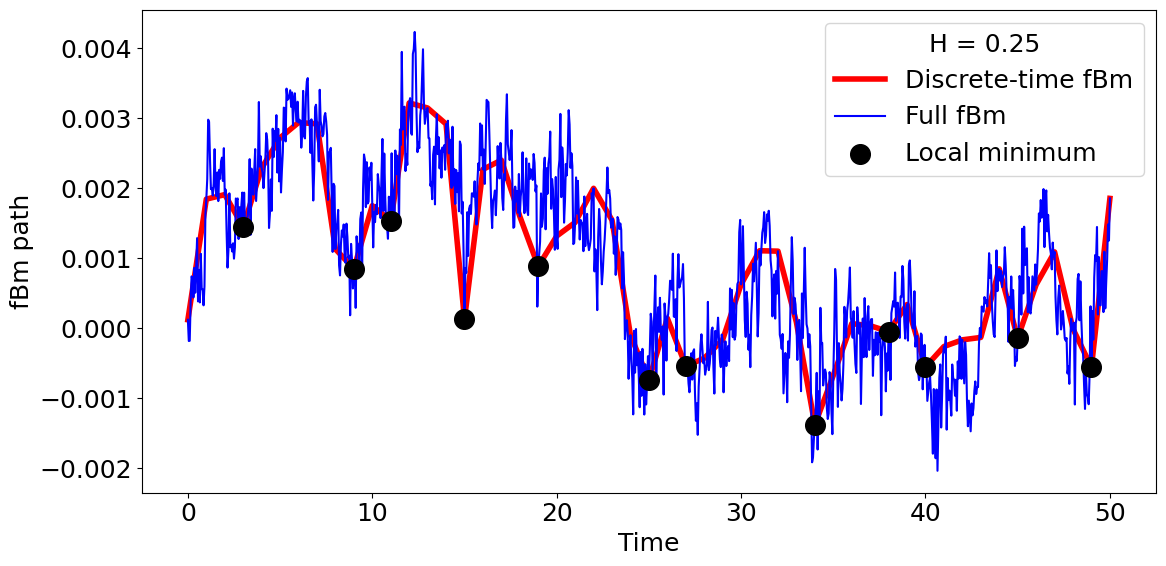}
    \includegraphics[scale=0.29]{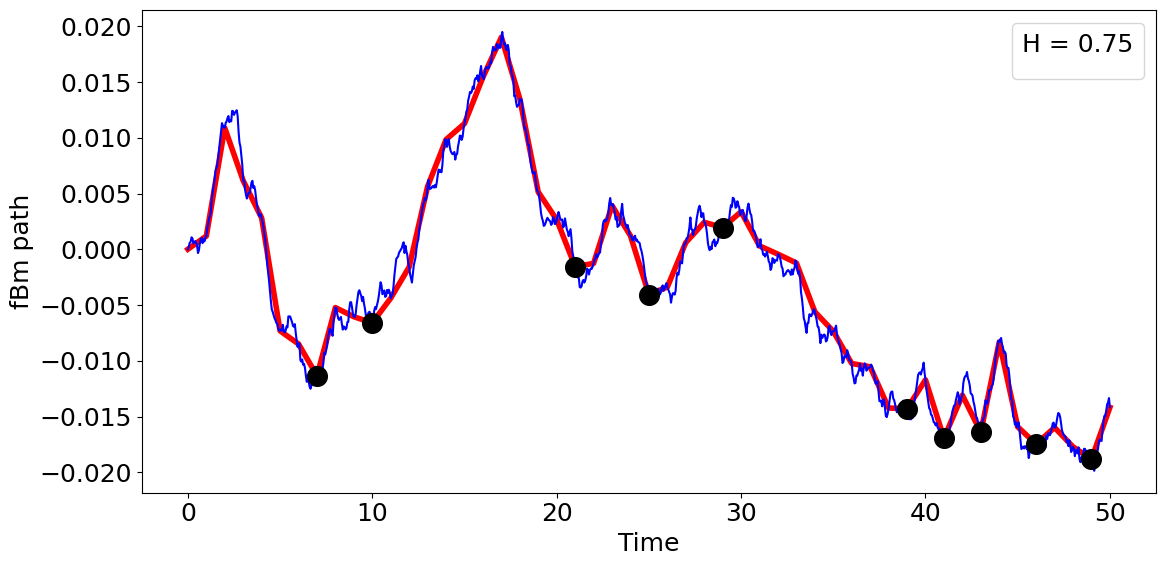}
    \caption{fBm trajectories of Hurst exponents $H=1/4$ and $3/4$ (blue line in top and bottom figures, respectively), sampled at integer-valued times (red line). Black dots indicate the local minima in the sampled trajectories, the number of which strongly depends on the Hurst exponent.}
    \label{fig:cartoon}
\end{figure}

Recently, Kundu, Majumdar, and Schehr derived the exact distribution of the number of local minima for a broad class of Markovian symmetric walks~\cite{kundu2024universal,kundu2025constrained}, which is expressed explicitly in terms of combinatorial coefficients and converges to a Gaussian distribution with mean $N/4$ and variance $N/16$ as the number of steps $N$ grows. However, many real-world systems are non-Markovian, typically due to interactions with possibly hidden degrees of freedom \cite{wei2000single,mason1997particle,krapf2019spectral,dolgushev2025evidence}. For example, in the context of the above-mentioned applications, heartbeat and electroencephalogram signals  \cite{ivanov1999multifractality,lopes2009fractal} as well as pricing options \cite{zhang2021pricing,sottinen2001fractional}, are known to exhibit memory. Describing such non-Markovian dynamics is typically a difficult task. 

One of the most fundamental approaches to non-Markovian dynamics is fractional Brownian motion (fBm), a symmetric Gaussian process with stationary increments that has long-range correlations \cite{Mandelbrot:1968}. While, strictly speaking, fBm is a continuous, non-smooth process, in real-world applications it is sampled in discrete time. The resulting discrete-time process $X_0,\dots,X_{N}$ (see Fig.~\ref{fig:cartoon}) has increments $\phi_i = X_{i+1} - X_{i}$, called fractional Gaussian noise (fGn), which follow a multivariate Gaussian distribution:
\begin{align}\label{incr_dist}
P_N(\Phi) &= \frac{1}{\sqrt{(2\pi)^N\det{\bf R}}} \exp \left( - \frac{1}{2} \Phi^T {\bf R}^{-1} \Phi \right),
\end{align}
where ${\Phi = (\phi_0, \dots, \phi_{N-1})}$, and ${{\bf R} = (\rho_{ij})}$ is the correlation matrix, whose elements depend only on the distance between increments, $\rho_{ij} = \rho_{0|i-j|}$ and   
\begin{align}\label{rho_k}
\rho_{0k} = \frac{1}{2}  \left( |k-1|^{2H} + |k+1|^{2H}-2k^{2H} \right),
\end{align}
where $H$ ($0<H<1$) is the Hurst exponent \footnote{It was shown in Ref.~\cite{taqqu1975weak} that the discrete-time fBm $X_N$ converges to the continuous fBm that is widely used for construction of fBm \cite{davies1987tests,wood1994simulation,dieker2004simulation,enriquez2004simple,huang2017extremal,walter2020sampling}.}.\nocite{taqqu1975weak,davies1987tests,wood1994simulation,dieker2004simulation,enriquez2004simple,huang2017extremal,walter2020sampling} The large $N$ behavior of the corresponding variance $ \text{Var}(X_N)$ of the position $X_N$ is given by  $\text{Var}(X_N)\propto N^{2H}$, so that, for $H<1/2$, fBm exhibits subdiffusive motion; for ${H>1/2}$, it is superdiffusive; and only for $H=1/2$ is the motion diffusive and Markovian, with $\rho_{0k}=0$.

fBm has broad applications across various fields. It has been shown to effectively model the subdiffusive behavior of telomeres within the cell nucleus \cite{burnecki2012universal,Stadler:2017,krapf2019spectral}, as well as the constrained motion of chromosomal loci \cite{weber2010bacterial,bronshtein2015loss}. The same framework has also proven useful for describing the intracellular transport of lipid granules during early mitosis \cite{jeon2011vivo}, and for characterizing the dynamics of beads suspended in viscoelastic media \cite{mason1997particle,ernst2012fractional,krapf2019spectral,dolgushev2025evidence}. Furthermore, fBm provides insight into the motion of tracer particles in crowded fluidic environments \cite{szymanski2009elucidating}, and notably captures the superdiffusive transport of vacuoles in amoeboid cells \cite{reverey2015superdiffusion,krapf2019spectral}. 

In the context of the minima number --- which falls within the field of extreme-value statistics --- several results have been obtained for fBm, including:  survival probability  \cite{Krug:1997,Molchan:1999,levernier2019survival,Monch:2022,levernier2022everlasting}, mean first-passage time in confinement \cite{Guerin:2016},  time of maximum \cite{Delorme:2015}, the statistics of record ages \cite{regnier2023record}, the splitting probability to reach a remote rather than a nearby target \cite{zoia2009asymptotic,wiese2019first,dolgushev2025evidence}, and the exploration dynamics of $d$-dimensional spaces \cite{regnier2024visitation}. Here we focus on the full characterisation of the minima number $m_N$ in fBm, which formally can be written as 
\begin{align}\label{min_def}
m_N &= \sum_{i=1}^{N-1} \Theta(-\phi_{i-1}) \Theta(\phi_i).
\end{align}
Here $\Theta(\phi)$ is the Heaviside function, so that the random variable $\Theta(-\phi_{i-1}) \Theta(\phi_i)$ represents change in the slope of $X_{i}$.

The mean of the number of local minima in fBm, studied in \cite{toroczkai2000extremal,huang2017extremal},
\begin{align}
\langle m_N \rangle = \frac{N-1}{4}\left[1 - \frac{2}{\pi} \arcsin \rho_{01}\right].\label{eq_mean} 
\end{align}
is linear for large $N$ as in Markovian case \cite{kundu2024universal}. Meanwhile its prefactor is modified, but it depends only on the nearest-neighbor increment correlation $\rho_{01}$. In fact, it is the same as for the so-called autoregressive model AR(1) \cite{box2015time,Scheffer2009}, which is a discretized Ornstein-Uhlenbeck process \cite{larralde2004statistical}, and for the model with the correlation matrix:
\begin{align}
    \rho_{ij}=
    \begin{cases}
    \rho_{01} & \text{ for } |i-j| = 1,\\
    0 & \text{ for } |i-j| > 1.\\
    \end{cases}\label{rho_AR}
\end{align}
Therefore, although the mean $\langle m_N \rangle$ deviates from the Markovian behavior (which is obtained by setting ${\rho_{01}=0}$), it does not incorporate the key feature of the fBm process --- the long-range correlations described by Eq.~\eqref{rho_k}. 

Conversely, the variance of the number of local minima, $\text{Var}(m_N)$, for fBm exhibits a remarkably rich behavior that accounts for all increment correlations $\{\rho_{0k}\}$. Depending on the value of the Hurst exponent $H$,
\begin{align}
\text{Var}(m_N)\sim & 
\begin{cases}
c_H N, & H<\frac{3}{4},\\
\frac{9(\sqrt{2}-1)}{64\pi^2} N\log N + b_{3/4}N, & H=\frac{3}{4},\\
\frac{4^{1-H}-1}{2\pi^2}\frac{H^2(2H-1)}{(4H-3)}N^{4H-2},  & H>\frac{3}{4}.\label{variance_fbm}
\end{cases}
\end{align}
 at leading order in $N$  (see Supplemental Material (SM) \footnote{See Supplemental Material at [url] for detailed calculations and numerical checks of asymptotics, which includes Refs.~\cite{toroczkai2000extremal,childs1967reduction,huang2017extremal,kundu2024universal,kundu2025constrained,box2015time,SinnKeller2011,PipirasTaqqu2017,DobrushinMajor1979,Taqqu1979,Slud1994,VeilletteTaqqu2013,leonenko2025numerical,BreuerMajor1983,GiraitisSurgailis1985,AzmoodehPeccatiPoly2016,maejima2007wiener}}\nocite{toroczkai2000extremal,childs1967reduction,huang2017extremal,kundu2024universal,kundu2025constrained,box2015time,SinnKeller2011,PipirasTaqqu2017,DobrushinMajor1979,Taqqu1979,Slud1994,VeilletteTaqqu2013,leonenko2025numerical,BreuerMajor1983,GiraitisSurgailis1985,AzmoodehPeccatiPoly2016,maejima2007wiener}  for detailed calculations and numerical checks of asymptotics, as well as for definitions of coefficients $c_H$ and $b_{3/4}$). We note that the asymptotic scaling exponents of Eq.~\eqref{variance_fbm} also follow from Ref.~\cite{SinnKeller2011}, which considers the variance of the number of zero crossings of fGn in a unit interval.  Because the fGn $\phi_i$ is a discrete derivative of the dicrete-time fBm $X_{i}$, $\phi_i = X_{i+1} - X_{i}$, the number of local minima $m_N$ of fBm is equal to the number of zero crossings of fGn in a unit interval multiplied by $N/2$ and their variances are related by the factor $N^2/4$, respectively. However, in the regime $H>3/4$, which is a key analytic result, Ref.~\cite{SinnKeller2011} contains an error in the prefactor; our result~\eqref{variance_fbm} corrects this prefactor (see SM for two independent proofs). Importantly, this behavior comes from the long-range increment correlations $\rho_{0k}$. Given that $\rho_{0k}\propto k^{2H-2}$ for $k\rightarrow\infty$ and that the correlation function $\rho_{0k}$ is \emph{squared} in $\text{Var}(m_N)\propto\sum_{k=2}^{N-2} (N-k-1)\rho_{0k}^2$, the correlations become non-summable when $2(2H-2)=-1$, i.e. at $H=3/4$, which is not the usual threshold value $H=1/2$ that separates fBm sub- and superdiffusion \footnote{On the level of a path-integral representation, the fBm action involves a kernel with decay exponent $2-2H$~\cite{benichou2024unifying}; here the correlations of the minima involve the squared correlations of fBm increments, which are expected to be controlled by an effective kernel with exponent $4-4H$ (which becomes marginal, equal to $1$, at $H=3/4$).}\nocite{benichou2024unifying}. 
 
 In the regime $H\leq3/4$, our work provides further information on $\mathrm{Var}(m_N)$ beyond that which can be obtained from Ref.~\cite{SinnKeller2011}: For $H<3/4$, the prefactor $c_H$ does not seem to admit a compact analytic expression, as also stated in \cite{SinnKeller2011}. Nevertheless, we derive in the SM its behavior close to $H=1/2$ by treating $\epsilon = H-1/2$ as a small parameter, and obtain $c_H = \frac{1}{16} + \chi_1 \epsilon + \chi_2 \epsilon^2 + \mathcal{O}(\epsilon^3)$,
with $\chi_1 = (3\ln 3 - 4\ln 2)/(4\pi) \approx 0.042$ and $\chi_2 \approx 0.069$. This shows that non-Markovian effects already enter $\mathrm{Var}(m_N)$ at first order in a perturbative expansion around Brownian motion. In the marginal case $H=3/4$, we further compute the prefactor of the subleading term and find $b_{3/4} \approx 0.0630$, which is more than ten times larger than the prefactor of the leading term ($\approx 0.00590$). As a result, the crossover to the asymptotic $N \log N$ behavior is extremely slow in practice (see SM).

\begin{figure}[t!]
    \centering
    \begin{overpic}[width=0.4\textwidth]{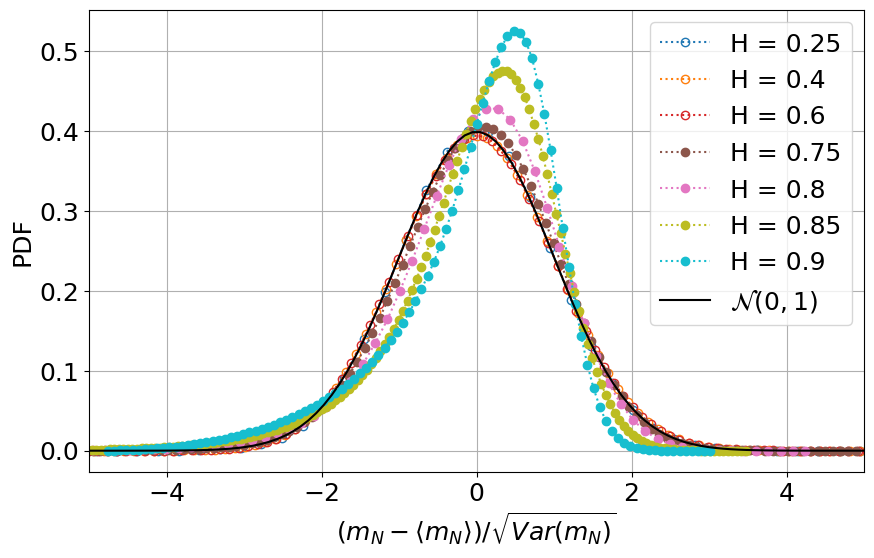}
        \put(0,63){\textbf{(a)}}
    \end{overpic}
    \begin{overpic}[width=0.4\textwidth]{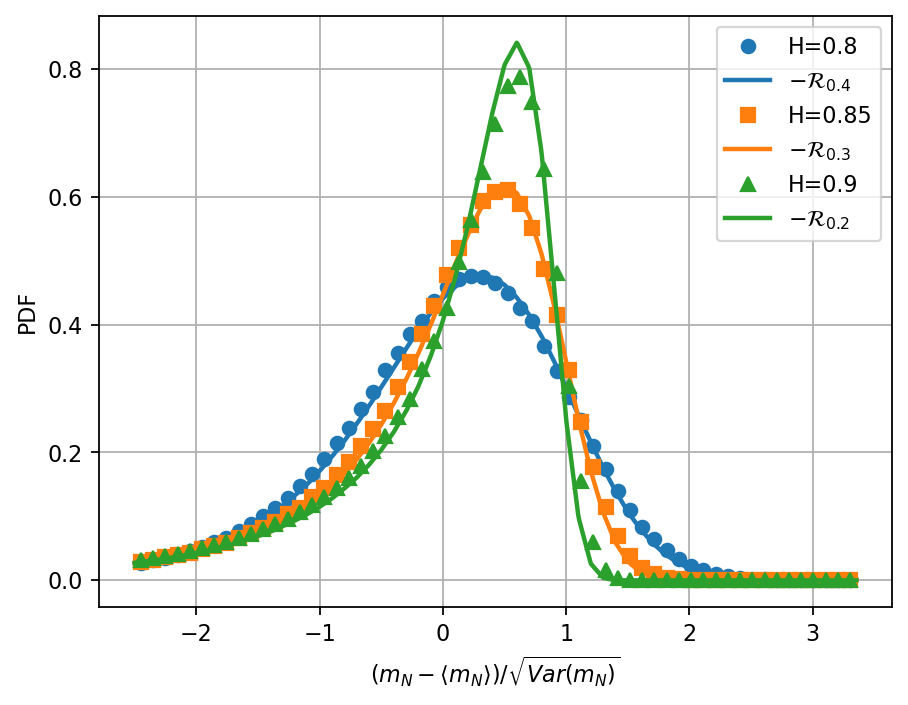}
        \put(0,73){\textbf{(b)}}
    \end{overpic}
    \caption{(a) Probability density function (PDF) of centered and normalized $m_N$ for ${N=1024}$, obtained from simulations (symbols), compared with Gaussian PDF (black line) and (b) for ${N=10^6}$ and various $H>3/4$ compared with the PDF of the negative Rosenblatt variable (lines).}
    \label{fig:PDF}
\end{figure}

The strong correlations of $\{\phi_i\}$  leading to the anomalously large (superlinear) fluctuations of $\text{Var}(m_N)$ have a fundamental consequence for the distribution of $m_N$, for which the central limit theorem (CLT) breaks down.  We show below this breakdown analytically and confirm it numerically through simulations (see Fig.~\ref{fig:PDF}(a)). In the analytic approach, we use the methods Ref.~\cite{PipirasTaqqu2017} for functionals of Gaussian correlated variables, which allow us to go beyond one-time distribution and to show the convergence of $m_N$ to the processes belonging to two different universality classes (Rosenblatt and Brownian motion), separated through the threshold value $H=3/4$. Note that, in the continuous setting, as outlined in the review by Kratz (Th.2.27) \cite{Kratz2006}, the limiting process of the related problem of zero crossings' number is given in the supercorrelated regime through a form involving implicitly the correlation function and is expressed through the multiple Wiener integral of the second order. This fundamental mathematical object, originally found by Slud \cite{Slud1994}, represents the Rosenblatt process, but practically is of little use. Here, in the discrete setting, we construct the Rosenblatt process in closed form with explicit analytic parametrisation.

First step consists in the expansion of $m_N$ in the probabilists’ Hermite polynomials $\mathrm{He}_n(\phi_i)$. Using
\begin{equation}
\Theta(\pm\phi_i)
= \frac{1}{2} \pm \sum_{\substack{n\ge1\\ n\ \mathrm{odd}}}
a_n\,\mathrm{He}_n(\phi_i),
\label{eq:HalfSpaceHermite_short}
\end{equation}
with $a_{2k+1}=(-1)^k/(\sqrt{2\pi}\,(2k+1)\,2^k\,k!)$ and $a_{2k}=0$, yields the explicit expansion of $M_i \equiv\Theta(-\phi_{i-1}) \,\Theta(\phi_i)$,
\begin{align}
M_i &=\frac14+\sum_{\text{odd }n}\frac{a_n}{2}\big(\mathrm{He}_n(\phi_{i})-\mathrm{He}_n(\phi_{i-1})\big)\label{eq:Mi-explicit_1}\\
&  -\sum_{\text{odd }m,n} a_m a_n\,\mathrm{He}_m(\phi_{i-1})\,\mathrm{He}_n(\phi_{i})
\label{eq:Mi-explicit_2}
\end{align}
The Hermite processes are classified according to the minimal nonzero order in their algebraic expansion in the Hermite polynomials \cite{PipirasTaqqu2017}. Equation~\eqref{eq:Mi-explicit_1} contains the linear terms proportional to $\mathrm{He}_1(\phi_{i})-\mathrm{He}_1(\phi_{i-1})$ of rank~$1$. However, in the number of minima $m_N$, Eq.~\eqref{min_def},
this contribution is a telescopic difference, leaving only the boundary increments $\phi_0$ and $\phi_{N-1}$ and thus does not contribute to $m_N$ asymptotically as $N\to\infty$. The term of Eq.~\eqref{eq:Mi-explicit_2} having minimal degree, with non-cancelled in $m_N$ increments $\{\phi_i\}$ for all $i$, is proportional to $\mathrm{He}_1(\phi_{i-1})\,\mathrm{He}_1(\phi_{i})$. This bilinear form can be expressed in terms of $\mathrm{He}_2(\dots)$, as we also show below. Therefore, the $m_N$ of Eq.~\eqref{min_def} has asymptotically ($N\to\infty$)  the Hermite rank $2$. In the regime of strong long-range correlations ($H>3/4$) leading to superlinear $\text{Var}(m_N)$, these quadratic terms will dominate all fluctuations (here we use Theorem~5.3.3 of Ref.~\cite{PipirasTaqqu2017} for general functionals of Gaussian correlated variables to  show that the higher order terms become irrelevant). In the opposite regime ($H<3/4$) all orders starting from order~$2$ matter, but despite of the long-range correlations, the Breuer-Major theorem \cite{BreuerMajor1983} for functionals of Gaussian correlated variables guarantees that the CLT holds.

\begin{figure*}
     \centering
    \begin{overpic}[width=0.32\textwidth]{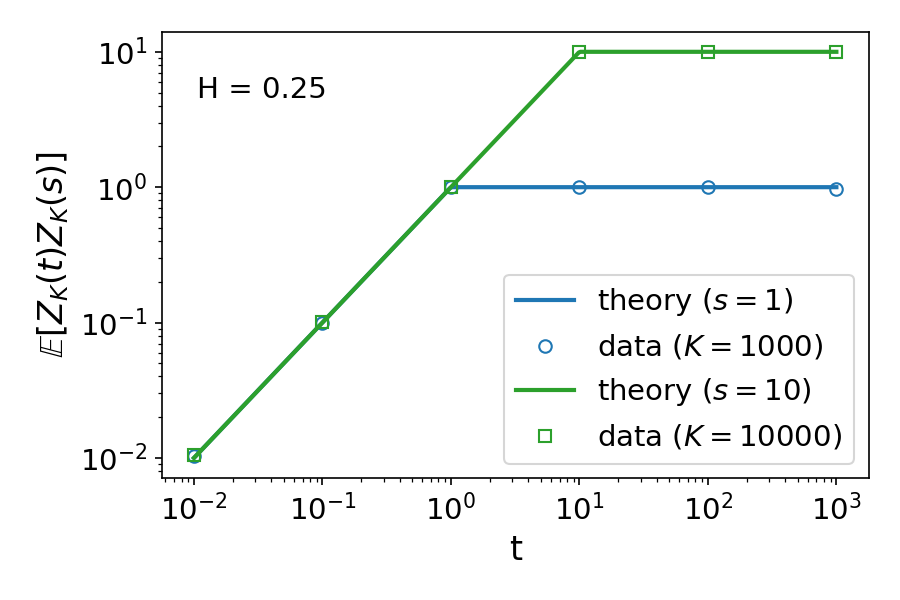}
        \put(0,60){\textbf{(a)}}
    \end{overpic}\hfill
    \begin{overpic}[width=0.32\textwidth]{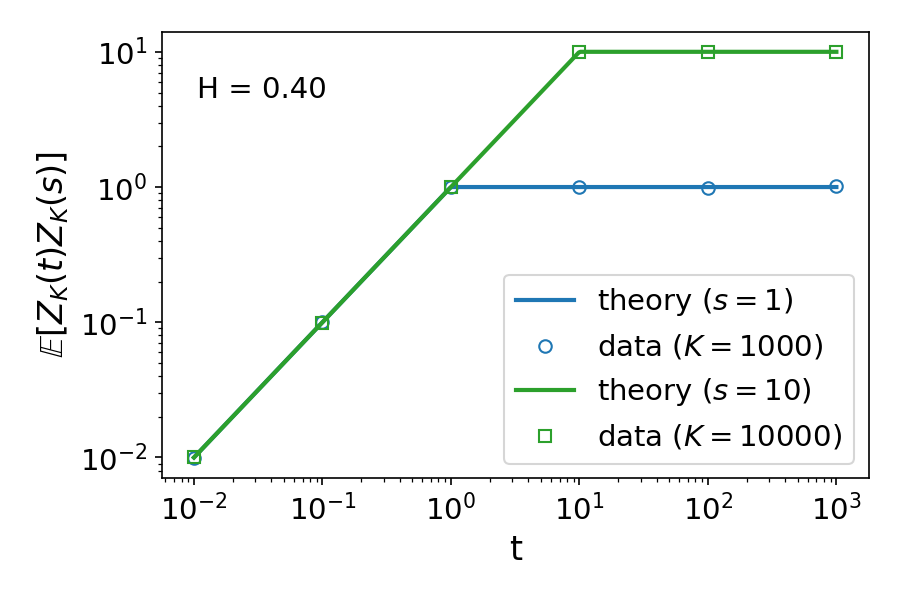}
        \put(0,60){\textbf{(b)}}
    \end{overpic}\hfill
    \begin{overpic}[width=0.32\textwidth]{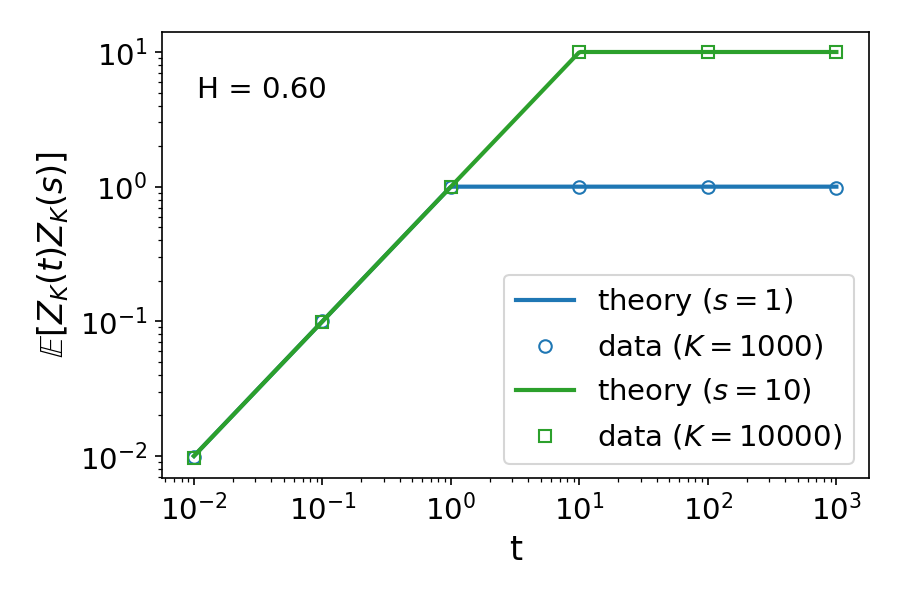}
        \put(0,60){\textbf{(c)}}
    \end{overpic}
    \vspace{0.5em}
    \begin{overpic}[width=0.32\textwidth]{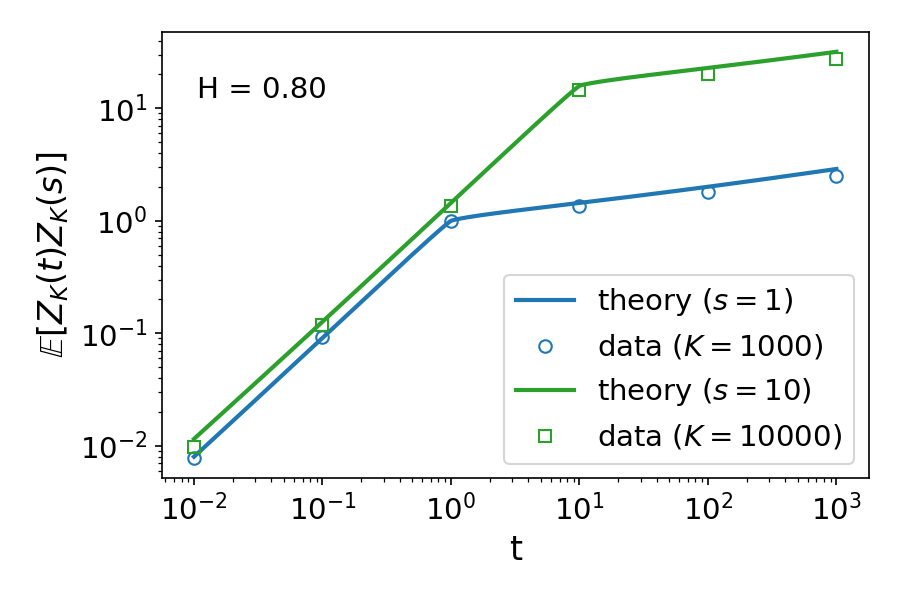}
        \put(0,60){\textbf{(d)}}
    \end{overpic}\hfill
    \begin{overpic}[width=0.32\textwidth]{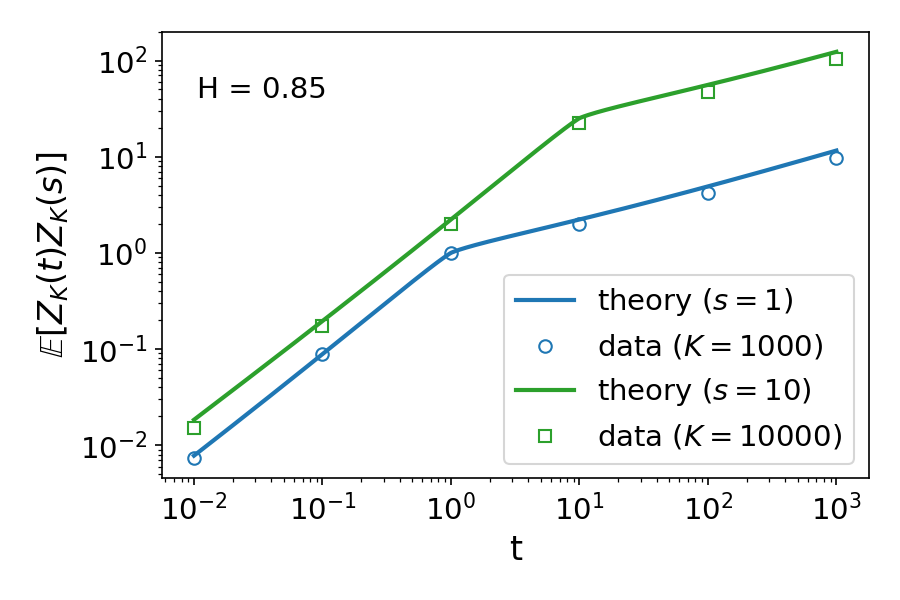}
        \put(0,60){\textbf{(e)}}
    \end{overpic}\hfill
    \begin{overpic}[width=0.32\textwidth]{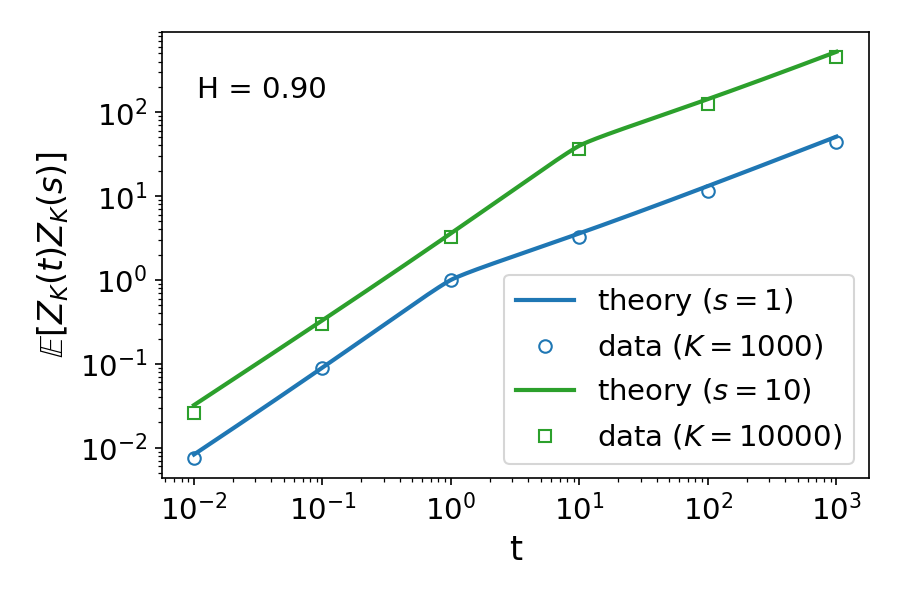}
        \put(0,60){\textbf{(f)}}
    \end{overpic}
    \caption{The covariance of the number of minima $\mathbb{E}[Z_K(t)Z_K(s)]$ (symbols) compared with theory (solid lines), For (a)-(c) the theoretical covariance is $\min(t,s)$ of the Brownian motion and for (d)-(f) is given by Eq.~\eqref{cov_ros} for the Rosenblatt process.  
}\label{fig:cov}
\end{figure*}

Because the rank~$2$ terms drive the CLT breaking, we are going now to analyse Eq.~\eqref{eq:Mi-explicit_2}, by spanning the quadratic part $M_i^\circ \equiv M_i - \langle M_i\rangle$ on the linearly independent vectors of the complete quadratic Wick basis 
\begin{align}
B_{1}\equiv\mathrm{He}_2(\phi_{i-1})+\mathrm{He}_2(\phi_i)=\phi_{i-1}^2+\phi_{i}^2-2,\nonumber\\ B_{2}\equiv\mathrm{He}_1(\phi_{i-1})\,\mathrm{He}_1(\phi_i)-\rho=\phi_{i-1}\phi_{i}-\rho,
\label{eq:Wick-basis}
\end{align}
where $\rho=\rho_{01}=\mathrm{Corr}(\phi_{i-1},\phi_i)=2^{2H-1}-1$. The Wick basis, Eq.~\eqref{eq:Wick-basis}, has the advantage that any centered quadratic functional of $(\phi_{i-1},\phi_i)$ can be written as a linear combination of $B_{1}$ and $B_{2}$, e.g., \begin{equation}
M_i^\circ = Q_i + R_i,\qquad
Q_i = c_1 B_1 + c_2 B_2,
\label{eq:Qi-def}
\end{equation}
where the coefficients $c_1,c_2$ are determined by the orthogonality conditions
\begin{equation}
\langle R_i,B_j\rangle = 0,\qquad j=1,2,\label{eq:projection}
\end{equation}
with $\langle A,B\rangle=\mathbb{E}[AB]$. In other words, $Q_i$ is the orthogonal projection of $M_i^\circ$ onto the
second order stochastic terms and the remainder $R_i$ collects all the components of order $q\neq 2$ (including the rank–1 term that telescopes and all
higher orders $q\ge3$). Solving the projection problem of Eq.~\eqref{eq:projection} yields (see SM for details):
\begin{equation}
Q_i
= \frac{\rho(\phi_{i-1}^2+\phi_{i}^2-2)
-2(\phi_{i-1}\phi_{i}-\rho)}{4\pi\sqrt{1-\rho^2}}.
\label{eq:Qi-final}
\end{equation}
 
Introducing the normalized sum and difference modes
\begin{equation}
V_i=\frac{\phi_i+\phi_{i-1}}{\sqrt{2(1+\rho)}},
\qquad
U_i=\frac{\phi_i-\phi_{i-1}}{\sqrt{2(1-\rho)}},
\end{equation}
which are standard and independent (for fixed $i$), we can rewrite the quadratic component as
\begin{equation}
Q_i = -\frac{\sqrt{1-\rho^2}}{4\pi}\left[\mathrm{He}_2(V_i)-\mathrm{He}_2(U_i)\right].
\label{eq:QVQU}
\end{equation}
Summing over $i$, we obtain the fluctuation decomposition
\begin{equation}
m_N-\langle m_N\rangle
= -\frac{\sqrt{1-\rho^2}}{4\pi}\sum_{i=1}^{N-1}\left[\mathrm{He}_2(V_i)-\mathrm{He}_2(U_i)\right]
+ \sum_{i=1}^{N-1}R_i.
\label{eq:fluct-decomp-Wick}
\end{equation}
The sum of the remainder $R_i$ contains (i) boundary increments that do not contribute asymptotically to ${m_N-\langle m_N\rangle}$ as $N\to\infty$ at any $H$ and 
(ii) the fluctuations of higher than quadratic order that are irrelevant for $H>3/4$ (Theorem~5.3.3 of Ref.~\cite{PipirasTaqqu2017}).

We now focus on the regime (ii), in which the high-$i$ terms linear in $\mathrm{He}_2(\dots)$ dominate the sum in Eq.~\eqref{eq:fluct-decomp-Wick}, leading to the superlinear $\mathrm{Var}(m_N)$ of Eq.~\eqref{variance_fbm}. Here, the $V$-mode inherits the long memory
\begin{align}\label{eq:cov_V}
    \mathrm{Cov}(V_1,V_{k+1})=\frac{2H(2H-1)}{1+\rho}\,k^{2H-2}+O(k^{2H-4}),
\end{align}
whereas the $U$-mode is two derivatives shorter, at $\mathrm{Cov}(U_1,U_{k})\propto k^{2H-4}$. Thus, the long-range dependence of ${m_N-\langle m_N\rangle}$ is carried entirely by the sum built on the terms $\{\mathrm{He}_2(V_i)\}$,
\begin{equation}
m_N-\langle m_N\rangle
\underset{N\rightarrow\infty}{\sim} -\frac{\sqrt{1-\rho^2}}{4\pi}\sum_{i=1}^{N-1}\mathrm{He}_2(V_i).
\label{eq:fluct-decomp-Wick-assympt}
\end{equation}
Equation~\eqref{eq:fluct-decomp-Wick-assympt} represents, up to negative constant prefactor, the celebrated form of the Rosenblatt random variable \cite{DobrushinMajor1979,Taqqu1979}, where the standard Gaussian variables $V_i$ are correlated, $\mathbb{E}(V_0V_k)\propto k^{-D}$ with $0<D<1/2$.
Using Eq.~\eqref{eq:cov_V}, we find that $D=2-2H$  
and according to the Dobrushin–Major–Taqqu theorem \cite{PipirasTaqqu2017,DobrushinMajor1979,Taqqu1979} we finally obtain the key exact result that, for $H>3/4$,
\begin{equation}
\frac{m_N-\langle m_N\rangle}{\sqrt{\mathrm{Var}(m_N)}} \Longrightarrow -\mathcal{R}_{D},
\label{eq:Rosenblatt-limit}
\end{equation}
where $\mathcal{R}_D$ is the canonical Rosenblatt random variable of unit variance. 

For $H\le 3/4$, Eq.~\eqref{eq:fluct-decomp-Wick} contains only short-range correlated random variables and the Breuer–Major type CLT applies \cite{PipirasTaqqu2017,BreuerMajor1983}: the fluctuations of $m_N$ are Gaussian \footnote{In the marginally non-summable case $H=3/4$ a CLT still holds (but with unusual normalization $\sqrt{N\log N}$), with logarithmically slow convergence to the normal distribution \cite{GiraitisSurgailis1985}.}\nocite{GiraitisSurgailis1985} and the centered and normalized $m_N$ converges to the Brownian motion $\mathcal{B}$ \cite{PipirasTaqqu2017,GiraitisSurgailis1985,AzmoodehPeccatiPoly2016},
\begin{equation}
\frac{m_N-\langle m_N\rangle}{\sqrt{\mathrm{Var}(m_N)}} \Longrightarrow \mathcal{B}.
\label{eq:Brownian-limit}
\end{equation}

Figure~\ref{fig:PDF} showing the one-time PDF of $m_N$ obtained from the fBm simulations \footnote{The fBm simulations are performed based on the Davies-Harte method \cite{davies1987tests,wood1994simulation,dieker2004simulation}, implemented in the fbm package in Python.}\nocite{davies1987tests,wood1994simulation,dieker2004simulation} confirms that $m_N$ follow either Gaussian or Rosenblatt distribution and with this the CLT-breaking at $H=3/4$ \footnote{We note that computation of the Rosenblatt PDF is an active topic in mathematical literature \cite{VeilletteTaqqu2013,leonenko2025numerical}; here we use the numerically obtained Rosenblatt PDF from Ref.~\cite{VeilletteTaqqu2013}.}.
 
Finally, Eqs.~\eqref{eq:Rosenblatt-limit} and \eqref{eq:Brownian-limit} represent the convergence in the process, which provide us the full statistics of $m_N$ at any number of times. Here we exemplify it on the case of two times by considering the covariance of
\begin{equation}
Z_K(t)=\frac{m_{[Kt]}-\langle m_{[Kt]}\rangle}{\sqrt{\mathrm{Var}(m_K)}},
\end{equation}
which is the basic tool widely used by physicists to describe ageing in the dynamics \cite{hofling2013anomalous}.

For $H>3/4$, $Z_K(t)$ converges to the
Rosenblatt process $\mathcal{R}(t)$, whose covariance reads \cite{maejima2007wiener}
\begin{equation}
\mathbb{E}[\mathcal{R}(t)\mathcal{R}(s)]
=\frac{1}{2}\big(t^{2(2H-1)}+s^{2(2H-1)}-|t-s|^{2(2H-1)}\big).\label{cov_ros}
\end{equation}
For $H\le3/4$, $\mathbb{E}[Z_K(t)Z_K(s)]$ converges to that of the standard Brownian motion $\mathbb{E}[\mathcal{B}(t)\mathcal{B}(s)]=\min(t,s)$ \cite{PipirasTaqqu2017}. Figure~\ref{fig:cov} confirms the two distinct behaviors of  $\mathbb{E}[Z_K(t)Z_K(s)]$ depending on $H$ and highlights the fundamental transition of the number of minima in fBm at $H>3/4$.

\textit{Acknowledgements.} The work was inspired by the talk of G. Schehr in Journées de Physique Statistique 2025 in Paris. We thank the anonymous referee for putting our attention on Refs.~\cite{SinnKeller2011,Slud1994,Kratz2006}.

\textit{Data availability.} The data that support the findings of this article are openly available~\cite{data}.

\end{document}


\title{SUPPLEMENTAL MATERIAL\\ Number of local minima in discrete-time fractional Brownian motion}
\author{Maxim Dolgushev}
\author{Olivier B\'enichou}
\date{\today}

\maketitle

\tableofcontents

\section{Number of local minima $m_N$ in a discrete process : direct calculation of the moments}\label{number}

We consider a discrete-time Gaussian process $X_0,\dots,X_{N}$ with stationary increments $\phi_i = X_{i+1} - X_{i}$. Following Toroczkai et al. \cite{toroczkai2000extremal}, the number of minima can be expressed as
\begin{align*}
m_N &= \sum_{i=1}^{N-1} \Theta(-\phi_{i-1}) \Theta(\phi_i)
\end{align*}
where $\Theta(\phi_i)$ is the Heaviside function. Thus, the calculation of the moments of $m_N$ reduces to the calculation of the correlation functions of $\Theta(\phi_i)$. 

\subsection{Mean of $m_N$}\label{sec:mean}
In particular, the first moment moment of $m_N$ for a process with stationary increments reads
\begin{align}
\langle m_N \rangle &= \sum_{i=1}^{N-1} \langle \Theta(-\phi_{i-1}) \Theta(\phi_i) \rangle = (N-1) \langle \Theta(-\phi_0) \Theta(\phi_1) \rangle. \label{first_mom} 
\end{align}
With this the first moment depends on a two-point correlation function only. To compute it, we adapt the method of Childs~\cite{childs1967reduction}, with difference that the arguments of the Heaviside functions now have opposite signs. Using the short notation $\Phi = (\phi_0, \dots, \phi_{N-1})$,
\begin{align}\label{twopoint_def}
\langle \Theta(-\phi_0) \Theta(\phi_1) \rangle &= \int_{\mathbb{R}^N}  \Theta(-\phi_0) \Theta(\phi_1) P_N(\Phi) d^N\Phi,
\end{align}
where the multivariate Gaussian distribution
\begin{align*}
P_N(\phi) &= \frac{1}{\sqrt{(2\pi)^N\det{\bf R}}} \exp \left( - \frac{1}{2} \Phi^T {\bf R}^{-1} \Phi \right) 
\end{align*}
is defined through the correlation matrix 
${\bf R} = (\rho_{ij})$. The integral \eqref{twopoint_def} can be calculated in the Fourier space. Defining the Fourier transform
\begin{align*}
    \mathcal{F}(f(\phi))\equiv \hat{f}(\omega)=\int_{-\infty}^{\infty}f(\phi)e^{-\mathrm{i}\phi\omega}d\phi
\end{align*}
and denoting by $\Omega=(\omega_0,\dots,\omega_{N-1})$, we have
\begin{align*}
\hat{P}_N(\Omega) &= \hat{P}_N(-\Omega) = \exp \left( - \frac{1}{2} \Omega^T {\bf R} \Omega \right).
\end{align*}
Furthermore,
\begin{align*}
\hat{\Theta}(\omega) &= \lim_{\epsilon \to 0} \left\{ \pi \delta(\omega) - \frac{i \omega}{\omega^2 + \epsilon^2} \right\} \\
\hat{1} &= 2 \pi \delta(\omega). 
\end{align*}
Recalling the Parseval's theorem
\[
\int_{-\infty}^{\infty} f(\phi) g(\phi) d\phi = \frac{1}{2\pi}\int_{-\infty}^{\infty} \hat{F}(\omega) \hat{G}(-\omega) d\omega
\]
and using a short-hand notation $\frac{i \omega}{\omega^2 + \epsilon^2}\equiv\frac{i }{\omega}$ we get
\begin{align}
\langle \Theta(-\phi_0) \Theta(\phi_1) \rangle &= \frac{1}{4\pi^2} \int_{\mathbb{R}^N} \left[ \pi \delta(\omega_0) + \frac{i }{\omega_0} \right] \left[ \pi \delta(\omega_1) - \frac{i }{\omega_1} \right] \prod_{j=2}^{N-1} \delta(\omega_j) \exp \left( - \frac{1}{2} \Omega^T {\bf R} \Omega \right) d^N \Omega \nonumber\\
&= \frac{1}{4}+\frac{1}{4\pi^2} \int_{\mathbb{R}^2}  \frac{d\omega_0 d\omega_1}{\omega_0\omega_1} e^{-\frac{1}{2} \left( \omega_0^2 + \omega_1^2 + 2\rho_{01} \omega_0 \omega_1  \right)}\nonumber\\
&= \frac{1}{4} - \frac{1}{2\pi} \arcsin (\rho_{01}).\label{eq_childs} 
\end{align}
Equation~\eqref{eq_childs} was obtained by Toroczkai et al.\cite{toroczkai2000extremal} and applied to fBm minima by Huang et al. \cite{huang2017extremal}. For $\rho_{01}=0$ we get readily the Markovian result \cite{kundu2024universal,kundu2025constrained}.

\subsection{Second moment of $m_N$}
We start with the determination of the second moment of $m_N$.
\begin{align*}
\langle m_N^2 \rangle &=  \sum_{i=1}^{N-1} \sum_{j=1}^{N-1} \langle\Theta(-\phi_{i-1}) \Theta(\phi_i) \Theta(-\phi_{j-1}) \Theta(\phi_j) \rangle.
\end{align*}
Using the increments' stationarity, we get
\begin{itemize}
    \item for $i = j$: $\langle \Theta^2(-\phi_{i-1}) \Theta^2(\phi_i) \rangle=\langle \Theta(-\phi_{i-1}) \Theta(\phi_i) \rangle =  \langle m_N \rangle/(N-1)$;
    \item for $|i-j|=1$: $\langle \Theta(-\phi_{i-1}) \Theta(\phi_i) \Theta(-\phi_{i}) \Theta(\phi_{i+1}) \rangle=0$;
    \item for $|i-j|=k\geq 2$:
\end{itemize}
\begin{align*}
\langle \Theta(-\phi_{i-1}) \Theta(\phi_i) \Theta(-\phi_{i+k-1}) \Theta(\phi_{i+k}) \rangle = \langle \Theta(-\phi_0) \Theta(\phi_1) \Theta(-\phi_k) \Theta(\phi_{k+1}) \rangle \equiv a_k
\end{align*}

This allows us to write
\begin{align}
\langle m_N^2 \rangle &= \langle m_N \rangle +  2 \sum_{k=2}^{N-2} (N-k-1) a_k.\label{second_mom}
\end{align}
Following the method of Ref.~\cite{childs1967reduction} described in Sec.~\ref{number}, we calculate the four-point correlation function $a_k$ in the Fourier space: 
\begin{align}
a_k &= \frac{1}{16 \pi^4} \int_{\mathbb{R}^4} \left[ \pi \delta(\omega_0) + \frac{i }{\omega_0} \right] \left[ \pi \delta(\omega_1) - \frac{i }{\omega_1} \right]\left[ \pi \delta(\omega_k) + \frac{i }{\omega_k} \right] \left[ \pi \delta(\omega_{k+1}) - \frac{i }{\omega_{k+1}} \right]e^{ - \frac{1}{2} \left[ \omega_0^2 + \omega_1^2 + \omega_k^2 + \omega_{k+1}^2\right]} \nonumber \\
&\quad \times \exp \left\{ - \left[ \rho_{01} \omega_0 \omega_1 + \rho_{0k} \omega_0 \omega_k+ \rho_{0k+1} \omega_0 \omega_{k+1} +\rho_{1k} \omega_1 \omega_k + \rho_{1k+1} \omega_1 \omega_{k+1}+ \rho_{kk+1} \omega_k \omega_{k+1} \right] \right\}d^4\omega \nonumber\\
&= \frac{1}{16} \left[1+ \frac{2}{\pi}\left(\arcsin\rho_{0k}+\arcsin\rho_{1k+1} -\arcsin\rho_{01}-\arcsin\rho_{0k+1}-\arcsin\rho_{1k}-\arcsin\rho_{kk+1} \right) +I_k \right].\label{a_k}
\end{align}
Here
\begin{align*}
I_{k} &= \frac{1}{\pi^4}\int_{\mathbb{R}^4}  \frac{d\omega_0d\omega_1d\omega_kd\omega_{k+1}}{\omega_0\omega_1\omega_k\omega_{k+1}}e^{ - \frac{1}{2} \left[ \omega_0^2 + \omega_1^2 + \omega_k^2 + \omega_{k+1}^2\right]}\\
&\quad \times \exp \left\{ - \left[ \rho_{01} \omega_0 \omega_1 + \rho_{0k} \omega_0 \omega_k+ \rho_{0k+1} \omega_0 \omega_{k+1} +\rho_{1k} \omega_1 \omega_k + \rho_{1k+1} \omega_1 \omega_{k+1}+ \rho_{kk+1} \omega_k \omega_{k+1} \right] \right\}
\end{align*}
We make a change of variables $\omega_i=\sqrt{2} y_i$ and substitute $e^{-y_0^2}=y_0^2\int_{1}^{\infty}dt e^{-ty_0^2}$:
\begin{align*}
I_{k} &= \frac{1}{\pi^4}\int_{1}^{\infty}dt\int_{\mathbb{R}^4}  \frac{y_0dy_0dy_1dy_kdy_{k+1}}{y_1y_ky_{k+1}}e^{ -ty_0^2 -y_1^2 -y_k^2 -y_{k+1}^2-2\sum_{ij}\rho_{ij}y_iy_j}
\end{align*}
Integration by parts with respect to $y_0$ gives
\begin{align*}
I_k &= - \frac{\sqrt{\pi}}{\pi^4} \int_1^\infty\frac{dt}{t^{3/2}}  \int_{\mathbb{R}^3} \frac{dy_1 \, dy_k \, dy_{k+1}}{y_1 y_k y_{k+1}} \, G(y_1, y_k, y_{k+1}) \exp \left\{ - K(y_1, y_k, y_{k+1}) + \frac{1}{t} \left[ G(y_1, y_k, y_{k+1}) \right]^2 \right\}, 
\end{align*}
where $K=y_1^2+y_k^2+y_{k+1}^2+2(\rho_{1k} y_1 y_k + \rho_{1k+1} y_1 y_{k+1}+ \rho_{kk+1} y_k y_{k+1})$ and $G=\rho_{01} y_1 + \rho_{0k} y_k+ \rho_{0k+1} y_{k+1}$. The structure of function $G$ suggests splitting of the integral, $I_k=I_k^{(1)}+I_k^{(k)}+I_k^{(k+1)}$, with
\begin{align*}
I_k^{(\alpha)} &= - \rho_{0\alpha}\frac{\sqrt{\pi}}{\pi^4} \int_1^\infty\frac{dt}{t^{3/2}}  \int_{-\infty}^{\infty} \frac{ dy_{\gamma} }{y_{\gamma}}\int_{-\infty}^{\infty} \frac{dy_{\beta} }{y_{\beta}}\int_{-\infty}^{\infty}dy_{\alpha} e^{-K+\frac{1}{t}G^2} 
\end{align*}
Integration with respect to $y_{\alpha}$ and then to $y_{\beta}$ and $y_{\gamma}$ leads to
\begin{align*}
I_k^{(\alpha)}= - \frac{\rho_{0\alpha}}{\pi^3} \int_1^\infty \frac{dt}{t\sqrt{t - \rho_{0\alpha}^2}}\int_{-\infty}^{\infty} \frac{ dy_{\gamma} }{y_{\gamma}}\int_{-\infty}^{\infty} \frac{dy_{\beta} }{y_{\beta}} \, e^{-\frac{f_{\alpha\beta}y_{\beta}^2+f_{\alpha\gamma}y_{\gamma}^2+2g_{\alpha\beta\gamma}y_{\beta}y_{\gamma}}{(1 - \rho_{0\alpha}^2/t)}} 
= \frac{2 \rho_{0\alpha}}{\pi^2} \int_1^\infty \frac{dt}{t\sqrt{t - \rho_{0\alpha}^2}} \arcsin \left( \frac{g_{\alpha\beta\gamma}}{\sqrt{f_{\alpha\beta}f_{\alpha\gamma}}} \right),
\end{align*}
replacing $t$ by $1/u^2$,
\begin{align}
I_k^{(\alpha)}= \frac{4 \rho_{0\alpha}}{\pi^2} \int_0^1 \frac{du}{\sqrt{1 - \rho_{0\alpha}^2u^2}} \arcsin \left( \frac{g_{\alpha\beta\gamma}}{\sqrt{f_{\alpha\beta}f_{\alpha\gamma}}} \right).\label{I_k_alpha}
\end{align}
Here
\begin{align*}
f_{\alpha\beta} &= \left( 1 - u^2 \rho_{0\alpha}^2 \right) \left( 1 - u^2 \rho_{0\beta}^2 \right) - \left( \rho_{\alpha\beta} - u^2 \rho_{0\alpha}\rho_{0\beta} \right)^2,\\
g_{\alpha\beta\gamma} &= \left( 1 - u^2 \rho_{0\alpha}^2 \right) \left( \rho_{\beta\gamma} - u^2 \rho_{0\beta}\rho_{0\gamma} \right) - \left( \rho_{\alpha\beta} - u^2 \rho_{0\alpha}\rho_{0\beta} \right)\left( \rho_{\alpha\gamma} - u^2 \rho_{0\alpha}\rho_{0\gamma} \right)
\end{align*}

\subsection{Variance of $m_N$}

For Markovian processes, the variance $\text{Var}(m_N)$ is proportional to the mean $\langle m_N\rangle$, given that the term of dominant order $\mathcal{O}(N^2)$ in $\langle m_N^2\rangle$ coincides with this of $\langle m_N\rangle^2$ \cite{kundu2024universal,kundu2025constrained}. This respectively follows from our formulas by setting $\rho_{ij}=0$. 

\subsubsection{Short-range correlated process}
Let us consider first a process with only nearest-neighbor correlations: 
\begin{align*}
    \rho_{ij}=
    \begin{cases}
    \rho_{01} & \text{ for } |i-j| = 1\\
    0 & \text{ for } |i-j| > 1.\\
    \end{cases}
\end{align*}
In this case $f_k=f_{k+1}=(1-u^2\rho_{01}^2)$ and $f_{kk+1}=\rho_{01}(1-u^2\rho_{01}^2)$ so that $I_k=I_k^{(1)}=\frac{4 \rho_{01}\arcsin\rho_{01}}{\pi^2} \int_0^1 \frac{du}{\sqrt{1 - \rho_{01}^2u^2}}$ and 
\begin{align}\label{a_k_short}
a_k = \frac{1}{16} \left[1-\frac{4}{\pi}\arcsin\rho_{01} +\frac{4}{\pi^2}\arcsin^2\rho_{01} \right].
\end{align}
Inserting $a_k$ into Eq.~\eqref{second_mom} and using Eqs.~\eqref{first_mom} and \eqref{eq_childs} for the mean $\langle m_N \rangle$,
\begin{align}
\text{Var}_{\text{R1}}(m_N)=  \langle m_N \rangle +  [2-3(N-1)]a_k=\frac{N+1}{16}+\frac{N-3}{4\pi}\arcsin\rho_{01}-\frac{3N-5}{4\pi^2}\arcsin^2\rho_{01}.\label{variance_short}
\end{align}
We note that the variance is proportional to the mean $\langle m_N \rangle$ and of order $\mathcal{O}(N)$.

\begin{figure}
    \captionsetup{justification=raggedright} 
    \centering
    \begin{subfigure}[t]{0.32\textwidth}
        \begin{overpic}[width=\linewidth]{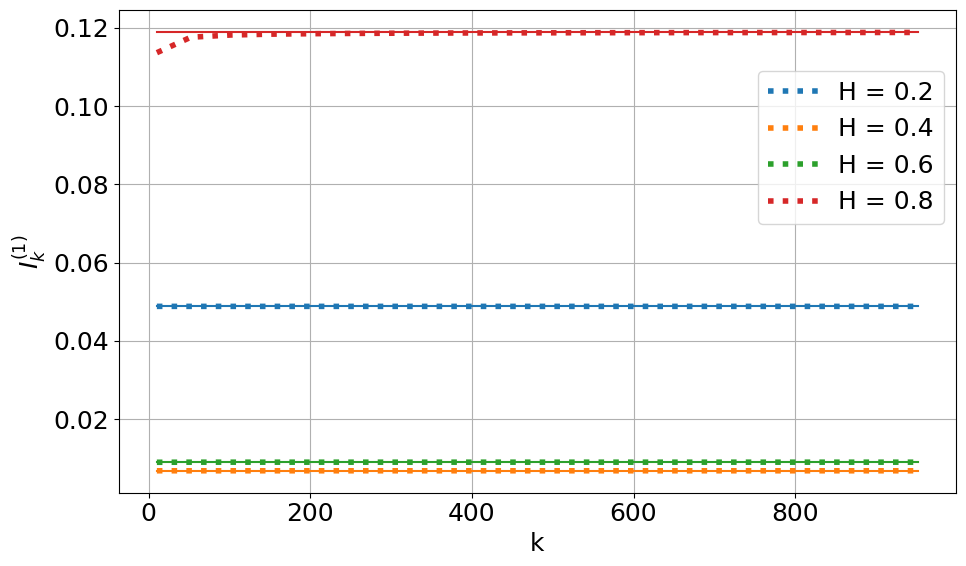}
            \put(0,100){\textbf{(a)}}
        \end{overpic}
    \end{subfigure}
    \hfill
    \begin{subfigure}[t]{0.32\textwidth}
        \begin{overpic}[width=\linewidth]{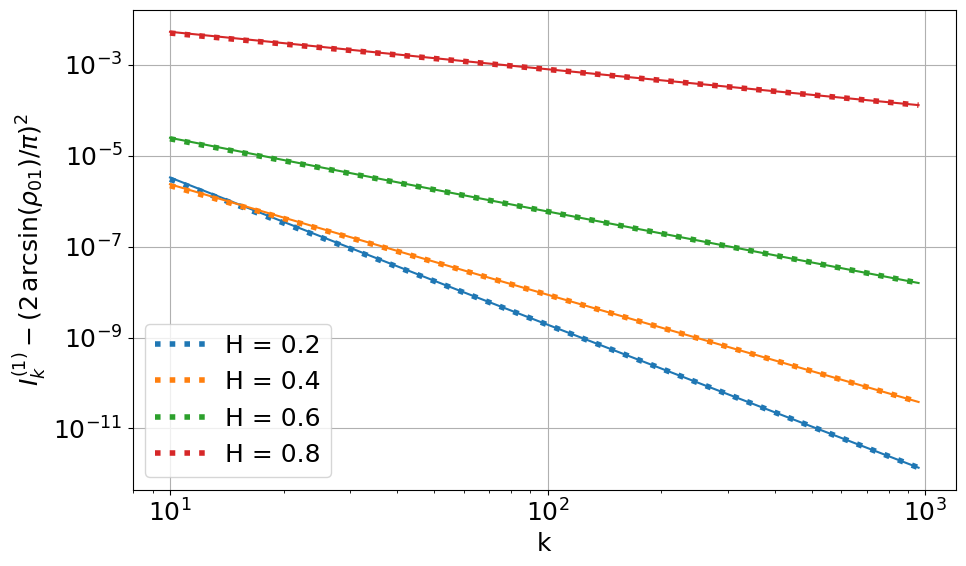}
             \put(0,100){\textbf{(b)}}
        \end{overpic}
    \end{subfigure}
    \hfill
    \begin{subfigure}[t]{0.32\textwidth}
        \begin{overpic}[width=\linewidth]{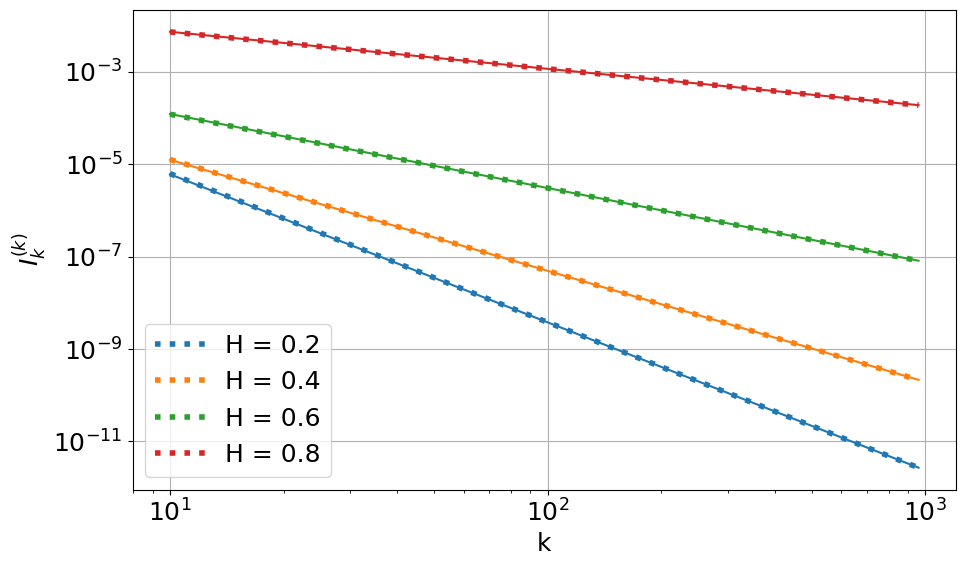}
             \put(0,100){\textbf{(c)}}
        \end{overpic}
    \end{subfigure}
\caption{Numerical calculation of the integrals $I_k^{(\alpha)}$ (dotted lines), Eq.~\eqref{I_k_alpha}, and comparison with the analytic Eqs.~\eqref{I_1}-\eqref{I_k} (solid lines). 
}\label{fig:int_I}
\end{figure}

\subsubsection{fBm}

For fBm of the Hurst exponent $H$ ($0<H<1$) the correlation matrix reads
\begin{align}\label{rho_ij}
\rho_{ij} = \rho(k = |i-j|) = \frac{\rho(0)}{2}  \left( |k-1|^{2H} + |k+1|^{2H}-2k^{2H} \right).
\end{align}
V.l.o.g. $\rho(0)=\sigma^2=1$. As we proceed to show, depending on the $H$-value, the variance $\text{Var}(m_N)$ is either ${\text{Var}(m_N)\propto N}$ for $H<3/4$ or  $\text{Var}(m_N)\propto N^{4H-2}$ for $H>3/4$ at leading order in $N$. 


We start the case of $H>3/4$ with the analysis of Eq.~\eqref{a_k}. There, the elements of the correlation matrix are $\rho_{kk+1}=\rho_{01}=2^{2H-1}-1$ and the other ones $\rho_{0k}=\rho_{1(k+1)}\sim\rho_{0(k+1)}\sim\rho_{1k}\sim H(2H-1)k^{2H-2}$ for $k\rightarrow\infty$. With this
\begin{align}\label{a_k_fbm}
a_k \underset{k\rightarrow\infty}{\sim}  \frac{1}{16} \left[1-\frac{4}{\pi}\arcsin\rho_{01} +I_k^{(1)}+I_k^{(k)}+I_k^{(k+1)} \right],
\end{align}
where $I_k^{(\alpha)}$ are given by Eq.~\eqref{I_k_alpha}. These integrals involve the functions:
\begin{align*}
    f_{1k}&=f_{k1}=\left( 1 - u^2 \rho_{01}^2 \right) \left( 1 - u^2 \rho_{0k}^2 \right) - \left( \rho_{1k} - u^2 \rho_{01}\rho_{0k} \right)^2\underset{k\rightarrow\infty}{\sim}1-u^2\rho_{01}^2+[(2\rho_{01}-1)u^2-1]\rho_{0k}^2\\
    f_{1(k+1)}&=f_{(k+1)1}=\left( 1 - u^2 \rho_{01}^2 \right) \left( 1 - u^2 \rho_{0(k+1)}^2 \right) - \left( \rho_{1(k+1)} - u^2 \rho_{01}\rho_{0(k+1)} \right)^2\underset{k\rightarrow\infty}{\sim}1-u^2\rho_{01}^2+[(2\rho_{01}-1)u^2-1]\rho_{0k}^2\\
    f_{k(k+1)}&=f_{(k+1)k}=\left( 1 - u^2 \rho_{0k}^2 \right) \left( 1 - u^2 \rho_{0(k+1)}^2 \right) - \left( \rho_{01} - u^2 \rho_{0k}\rho_{0(k+1)} \right)^2\underset{k\rightarrow\infty}{\sim}1-\rho_{01}^2\\
    g_{1k(k+1)}&=\left( 1 - u^2 \rho_{01}^2 \right) \left( \rho_{01} - u^2  \rho_{0k}\rho_{0(k+1)} \right) - \left( \rho_{1k} - u^2 \rho_{01}\rho_{0k} \right)\left( \rho_{1(k+1)} - u^2 \rho_{01}\rho_{0(k+1)}\right)\\
    &\underset{k\rightarrow\infty}{\sim}\rho_{01}(1-u^2\rho_{01}^2)+[(2\rho_{01}-1)u^2-1]\rho_{0k}^2\\
    g_{k1(k+1)}&=\left( 1 - u^2 \rho_{0k}^2 \right) \left( \rho_{1(k+1)} - u^2  \rho_{01}\rho_{0(k+1)} \right) - \left( \rho_{1k} - u^2 \rho_{01}\rho_{0k} \right)\left( \rho_{01} - u^2 \rho_{0k}\rho_{0(k+1)}\right)\underset{k\rightarrow\infty}{\sim}\rho_{0k}(1-\rho_{01})(1-u^2\rho_{01})\\
    g_{(k+1)1k}&=\left( 1 - u^2 \rho_{0(k+1)}^2 \right) \left( \rho_{1k} - u^2  \rho_{01}\rho_{0k} \right) - \left( \rho_{1(k+1)} - u^2 \rho_{01}\rho_{0(k+1)} \right)\left( \rho_{01} - u^2 \rho_{0k}\rho_{0(k+1)}\right)\underset{k\rightarrow\infty}{\sim}\rho_{0k}(1-\rho_{01})(1-u^2\rho_{01})
\end{align*}

\begin{figure}
    \captionsetup{justification=raggedright} 
    \centering
    \begin{subfigure}[t]{0.48\textwidth}
        \begin{overpic}[width=\linewidth]{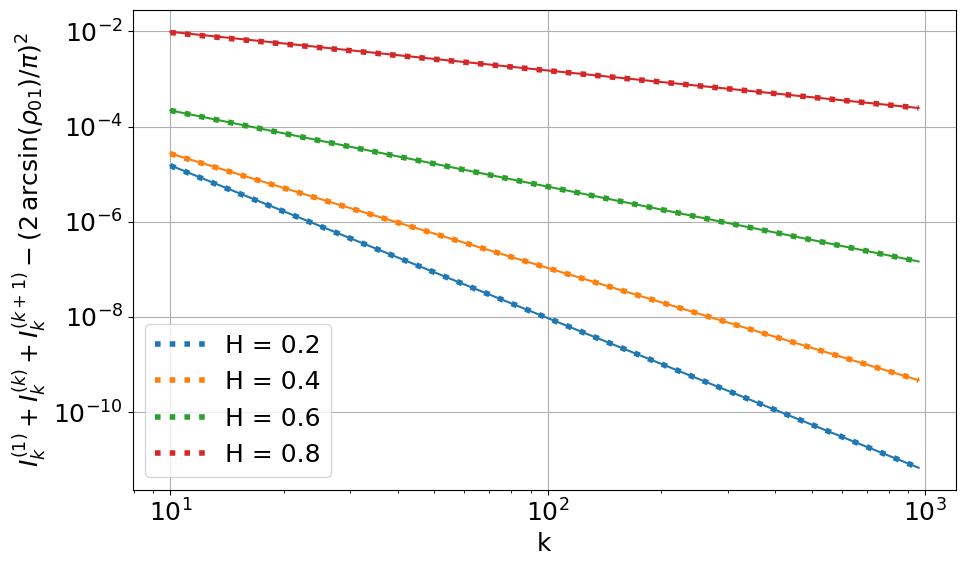}
            \put(0,150){\textbf{(a)}}
        \end{overpic}
    \end{subfigure}
    \hfill
    \begin{subfigure}[t]{0.48\textwidth}
        \begin{overpic}[width=\linewidth]{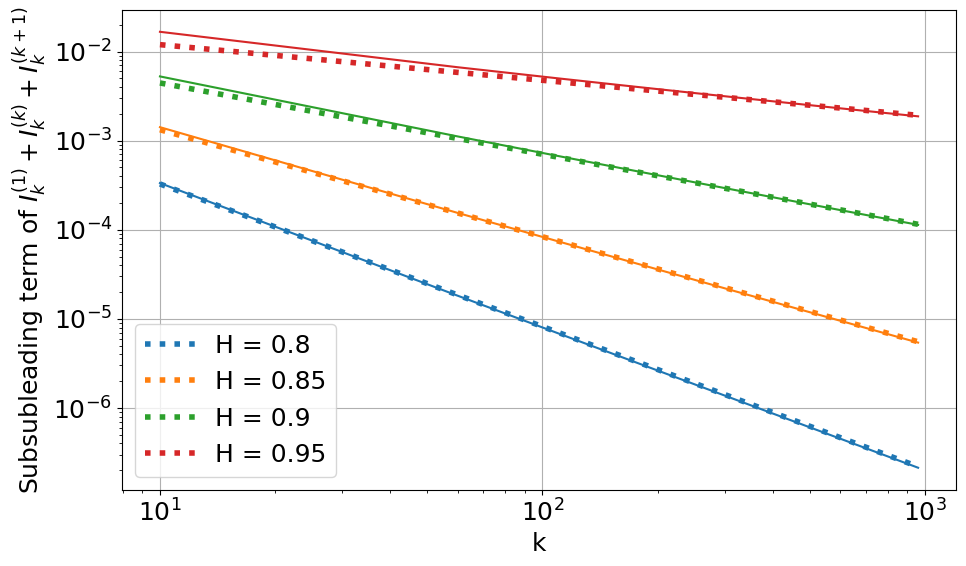}
             \put(0,150){\textbf{(b)}}
        \end{overpic}
    \end{subfigure}
\caption{Numerical check of analytic Eq.~\eqref{I_k_all} for $I_k^{(1)}+I_k^{(k)}+I_k^{(k+1)}$. In (a) solid lines show the subleading term of Eq.~\eqref{I_k_all} and in (b) the lines indicate the subsubleading decay $\mathcal{O}(k^{8H-8})$.
}\label{fig:int_I_all}
\end{figure}

Inserting these expressions into Eq.~\eqref{I_k_alpha},
\begin{align}
   I_k^{(1)}&\underset{k\rightarrow\infty}{\sim} \frac{4 \rho_{01}}{\pi^2} \int_0^1 \frac{du}{\sqrt{1 - \rho_{01}^2u^2}} \arcsin  (\rho_{01})+ \frac{4 \rho_{0k}^2\rho_{01}}{\pi^2}\left(\frac{1-\rho_{01}}{1+\rho_{01}}\right)^{\frac{1}{2}}\int_0^1 \frac{((2\rho_{01}-1)u^2-1)du}{(1 - \rho_{01}^2u^2)^{3/2}}\nonumber\\
   &= \frac{4}{\pi^2} \left(\arcsin \rho_{01} \right)^2-\frac{4\rho_{0k}^2}{\pi^2}\left[\frac{(1-\rho_{01})^2}{\rho_{01}(1+\rho_{01})}+\left(\frac{1-\rho_{01}}{1+\rho_{01}}\right)^{\frac{1}{2}}\frac{2\rho_{01}-1}{\rho_{01}^2}\arcsin \rho_{01}\right]\label{I_1}\\
    I_k^{(k)}&\underset{k\rightarrow\infty}{\sim}I_k^{(k+1)}\underset{k\rightarrow\infty}{\sim} \frac{4 \rho_{0k}^2}{\pi^2}\left(\frac{1-\rho_{01}}{1+\rho_{01}}\right)^{\frac{1}{2}}\int_0^1 \frac{(1 - \rho_{01}u^2)du}{\sqrt{1 - \rho_{01}^2u^2}} = \frac{2\rho_{0k}^2}{\pi^2}\left[\frac{1-\rho_{01}}{\rho_{01}}+\left(\frac{1-\rho_{01}}{1+\rho_{01}}\right)^{\frac{1}{2}}\frac{2\rho_{01}-1}{\rho_{01}^2}\arcsin \rho_{01}\right].\label{I_k}
\end{align}
Altogether, 
\begin{align}
   I_k^{(1)}+I_k^{(k)}+I_k^{(k+1)}\underset{k\rightarrow\infty}{\sim} \frac{4}{\pi^2} \left(\arcsin \rho_{01} \right)^2+\frac{8\rho_{0k}^2}{\pi^2}\frac{1-\rho_{01}}{1+\rho_{01}} +\mathcal{O}(k^{8H-8}).\label{I_k_all}
\end{align}
Combining this result with $a_k$ of Eq.~\eqref{a_k_fbm} and Eq.~\eqref{second_mom}, the terms of order $N^2$ cancel in the variance $\text{Var}(m_N)$. The remaining terms of leading order $N^{4H-2}$ come from the terms of $I_k^{(\alpha)}$ proportional to $k^{4H-4}$,
\begin{align}
\text{Var}(m_N)\underset{N\rightarrow\infty}{\sim}   \left(\frac{H(2H-1)}{\pi}\right)^2 \frac{1-\rho_{01}}{1+\rho_{01}} \sum_{k=2}^{N-2} (N-k-1)k^{4H-4}\underset{N\rightarrow\infty}{\sim}   \frac{4^{1-H}-1}{2\pi^2}\frac{H^2(2H-1)}{(4H-3)}N^{4H-2}.
\label{variance_fbm_fast}
\end{align}
We note that for $H>7/8$, the next order terms of the variance come from the next-order terms in $I_k^{(\alpha)}$ proportional to $k^{8H-8}$ (see Fig.~\ref{fig:int_I_all}(b)), giving a correction of $\mathcal{O}(N^{8H-6})$ for the variance; for $3/4<H<7/8$ the leading correction is of $\mathcal{O}(N)$. Eq.~\eqref{variance_fbm_fast} corrects the result on the zero crossing number by fractional Gaussian noise (Theorem 6.1(iii) of Ref.~\cite{SinnKeller2011}), for which the variance is related to $\text{Var}(m_N)$ through the coefficient $4/N^2$.

For $H=3/4$ the summation in Eq.~\eqref{variance_fbm_fast} gives logarithmic terms and the variance behaves as
\begin{align}
    \text{Var}(m_N)\underset{N\rightarrow\infty}{\sim}   \frac{9(\sqrt{2}-1)}{64\pi^2} N\log N + b_{3/4}N , \label{variance_fbm_marg}
\end{align}
where the coefficient $b_{3/4}\approx0.0630$ is computed based on Eq.~\eqref{second_mom}. Eq.~\eqref{variance_fbm_marg} compliments the result of Ref.~\cite{SinnKeller2011} with the important subleading order.   

\begin{figure}
   \captionsetup{justification=raggedright} 
    \centering
    \begin{subfigure}[t]{0.48\textwidth}
        \begin{overpic}[width=\linewidth]{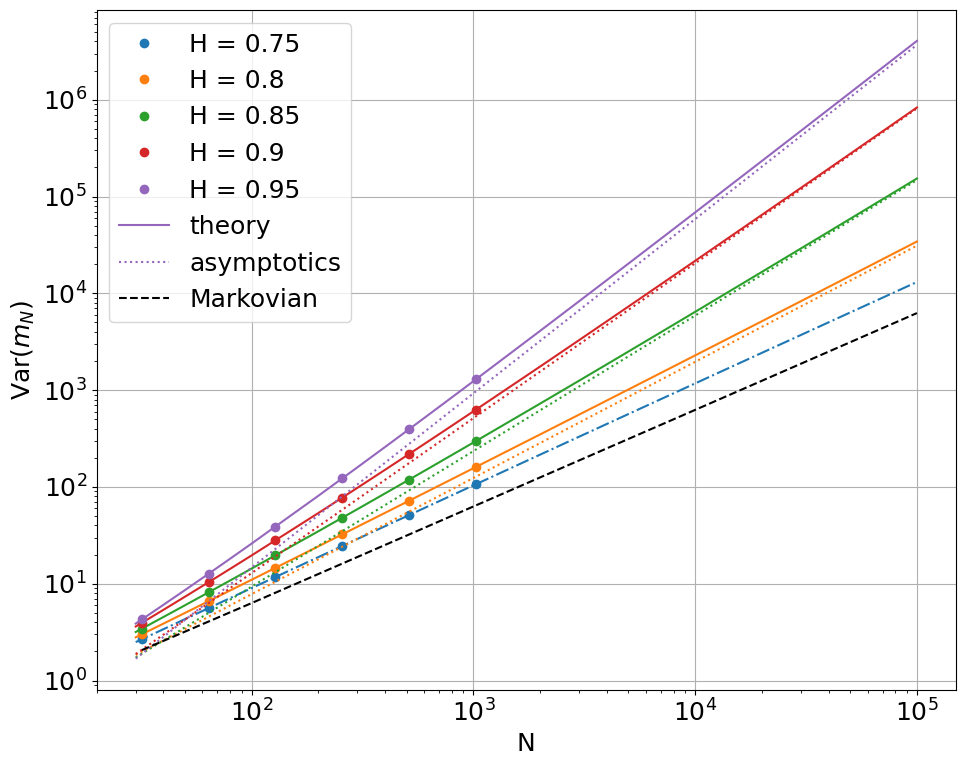}
            \put(0,170){\textbf{(a)}}
        \end{overpic}
    \end{subfigure}
    \hfill
    \begin{subfigure}[t]{0.48\textwidth}
        \begin{overpic}[width=\linewidth]{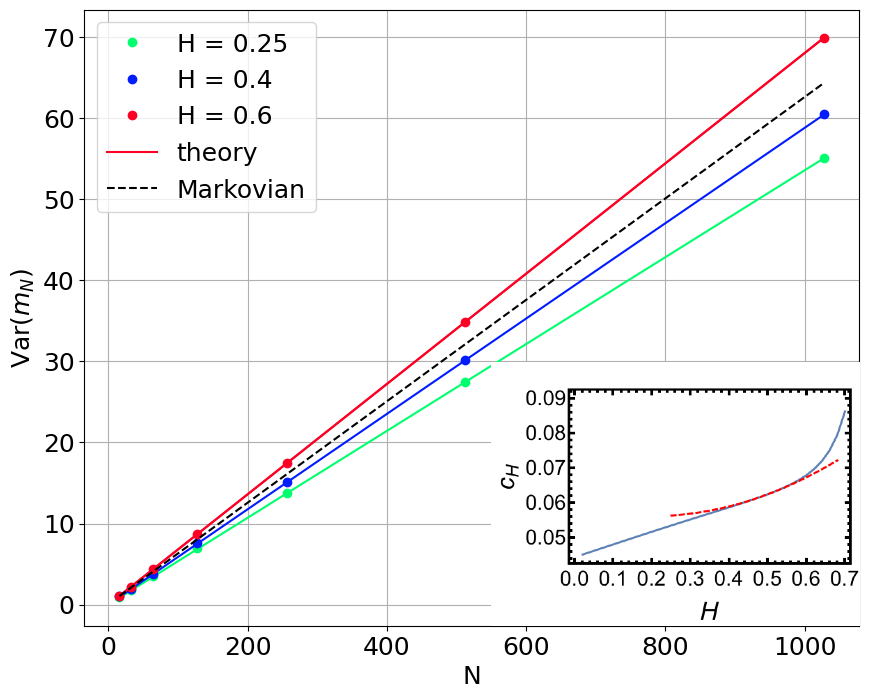}
             \put(0,170){\textbf{(b)}}
        \end{overpic}
    \end{subfigure}
    \caption{(a) Variance $\text{Var}(m_N)$ for fBm with $H>3/4$ calculated using Eqs.~\eqref{second_mom} and \eqref{a_k_fbm} (solid lines) is shown in comparison with simulations (symbols) and the asymptotic expression Eq.~\eqref{variance_fbm_fast} (dotted lines). For $H=3/4$ the blue dot-dashed curve represents Eq.~\eqref{variance_fbm_marg}. The black dashed line indicates the Markovian variance $(N+1)/16$. (b)~Variance $\text{Var}(m_N)$ for fBm with $H<3/4$ calculated using Eqs.~\eqref{second_mom} and \eqref{a_k_fbm} (solid lines) is shown in comparison with simulations (symbols). The black dashed line indicates the Markovian variance $(N+1)/16$. The variance is linear in $N$ for all values of $H<3/4$, $\text{Var}(m_N)\sim c_HN$. The blue solid line in the inset shows the coefficient $c_H$, Eq.~\eqref{c_H}, and the red dashed line shows its perturbative behavior around $H=1/2$, Eq.~\eqref{c_H_pert_final}.}
    \label{fig:var}
\end{figure}

For $H<3/4$, $4H-2<1$ and $\text{Var}(m_N)\propto N$ at leading order. Moreover, in contrast to the case of $H\geq 3/4$, the terms of order $N$ are dominated by the low values of $k$ in the sum of Eq.~\eqref{second_mom} and 
\begin{align}\label{c_H}
   \text{Var}(m_N)\underset{N\rightarrow\infty}{\sim}c_HN.
\end{align}
The asymptotics of Eqs.~\eqref{variance_fbm_fast}-\eqref{c_H} are tested in Fig.~\ref{fig:var} against full calculations based on Eq.~\eqref{second_mom} with $a_k$ from Eq.~\eqref{a_k_fbm} and numerical simulations. As can be inferred from the figure, the convergence for $H\geq3/4$ is quite slow.

Now, the explicit dependence of $c_H$ of Eq.~\eqref{c_H} on $H$ is unknown, but one can calculate it perturbatively around Brownian motion by setting $\epsilon=(H-1/2)$. Here we consider the variance at second order in $\epsilon$: 
\begin{align}
\text{Var}(m_N)=\text{Var}_{\text{R1}}(m_N)+A+B+\mathcal{O}(\epsilon^3),\label{variance_pert_def}
\end{align}
where $\text{Var}_{\text{R1}}(m_N)$ is given by Eq.~\eqref{variance_short}. With $\rho_{01}= 2\epsilon\ln2 +2\epsilon^2\ln^2 2 +\mathcal{O}(\epsilon^3)$,
\begin{align}
\text{Var}_{\text{R1}}(m_N)\underset{N\rightarrow\infty}{\sim}\left[\frac{1}{16}+\frac{\ln 2}{2\pi}\epsilon+(2\pi-12)\left(\frac{\ln 2}{2\pi}\right)^2\epsilon^2+\mathcal{O}(\epsilon^3)\right]N.
\end{align}
Next term of $\text{Var}(m_N)$ is evaluated based on the stationarity of fBm increments, $\rho_{ij} = \rho(k = |i-j|)$, 
\begin{align}
A= \frac{1}{4\pi} \sum_{k=2}^{N-2} (N-k-1)\left[2\arcsin\rho_{0k} -\arcsin\rho_{0k+1}-\arcsin\rho_{1k}\right]\underset{N\rightarrow\infty}{\sim} \frac{N}{4\pi}(\rho_{02}-\rho_{01})+\mathcal{O}(\epsilon^3).\label{A_pert_def}
\end{align}
Using  $\rho_{02}= \left( 3\ln3 - 4\ln2 \right) \epsilon +\left( 3\ln^2 3 - 4\ln^2 2 \right) \epsilon^2 +\mathcal{O}(\epsilon^3)$, we obtain
\begin{align}
A\underset{N\rightarrow\infty}{\sim}\frac{3}{4\pi}\left[(\ln 3-2\ln 2)\epsilon+(\ln^2 3-2\ln^2 2)\epsilon^2+\mathcal{O}(\epsilon^3)\right]N.\label{A_pert_final}
\end{align}
The remaining in Eq.~\eqref{variance_pert_def} term $B$ involves the elements of the correlation matrix $\rho_{ij}$ for $k = |i-j|>2$, which are at the second order in $\epsilon=(H-1/2)$ given by
\begin{align*}
\rho_{ij} = \left[ (k - 1)\ln(k - 1) + (k + 1)\ln(k + 1) - 2k\ln(k) \right]\epsilon + \left[ (k - 1)\ln^2(k - 1) + (k + 1)\ln^2(k + 1) - 2k\ln^2(k) \right]\epsilon^2+\mathcal{O}(\epsilon^3).
\end{align*}
With $I_k^{(\alpha)}$ defined in Eq.~\eqref{I_k_alpha}, which involve $f_{ij}=1+\mathcal{O}(\epsilon^2)$, $g_{k1(k+1)}=\rho_{0k}+\mathcal{O}(\epsilon^2)$ and $ g_{(k+1)1k}=\rho_{0k-1}+\mathcal{O}(\epsilon^2)$, leading to $I_k^{(\alpha)}= 4g_{\alpha\beta\gamma}\rho_{0\alpha}/\pi^2+\mathcal{O}(\epsilon^3)$,
\begin{align}
B= \frac{1}{8} \sum_{k=2}^{N-2} (N-k-1)\left[I_k^{(k)}+I_k^{(k+1)}\right]\underset{N\rightarrow\infty}{\sim} \frac{N}{2\pi^2}\sum_{k=2}^{N-2}(\rho_{0k}^2-\rho_{0k-1}\rho_{0k+1})+\mathcal{O}(\epsilon^3).\label{B_pert_def}
\end{align}
Finally,
\begin{align}
B= \underset{N\rightarrow\infty}{\sim} \frac{N}{2\pi^2}\left[ 9(\ln 3-2\ln 2)^2+r_{\infty} \right]\epsilon^2+\mathcal{O}(\epsilon^3).\label{B_pert_2}
\end{align}
where, using the notation $L(k,a)=k(k+a)\ln k \ln(k+a)$,
\begin{align*}
r_{\infty}=\lim\limits_{N \to \infty}\sum_{k=3}^{N-2}&\left[L(k-2,2)-2 L(k-2,3)+L(k-2,4)+L(k-1,0)-6 L(k-1,1)+6L(k-1,2)\right.\\
&\left.-2 L(k-1,3)+5L(k,0)-6 L(k,1)+L(k,2)+L(k+1,0)\right]\approx 0.829.
\end{align*}

Putting all terms into Eq.~\eqref{variance_pert_def}, the final result reads $\text{Var}(m_N)\underset{N\rightarrow\infty}{\sim}c_H N$, where
\begin{align}\label{c_H_pert_final}
    c_H=1/16+\chi_1\epsilon+\chi_2\epsilon^2+\mathcal{O}(\epsilon^3)
\end{align}
with $\chi_1=\rho_{02}^{(1)}/4\pi=(3\ln3-4\ln2)/4\pi\approx0.042$ (here $\rho_{02}^{(1)}=\lim\limits_{\epsilon \to 0}(\rho_{02}/\epsilon)$) and $\chi_2\approx0.069$.  

\newpage

\section{Hermite/Wick decomposition of $m_N$: Limit laws and covariances}

In this Section, we are going to analyze the number of minima
\begin{align}\label{m_N_def}
m_N &= \sum_{i=1}^{N-1} \Theta(-\phi_{i-1}) \Theta(\phi_i),
\end{align}
which is a functional of correlated Gaussian variables $\{\phi_i\}$, employing Hermite decomposition methods \cite{PipirasTaqqu2017}.

We use the probabilists' Hermite polynomials $\mathrm{He}_0(x)=1$, $\mathrm{He}_1(x)=x$, $\mathrm{He}_2(x)=x^2-1$, etc., with orthogonality $\mathbb{E}[\mathrm{He}_m(X)\mathrm{He}_n(X)]=\int_{-\infty}^{\infty} \mathrm{He}_m(x)\,\mathrm{He}_n(x)\,\frac{e^{-x^{2}/2}}{\sqrt{2\pi}}\,dx
= n!\,\delta_{nm}$
to expand the Heaviside function $\Theta(\phi_i)$:
\begin{align}
  \Theta(\pm\phi_i)=\frac{1}{2} \pm \sum_{\substack{n\ge1\\ n\ \mathrm{odd}}} a_n\,\mathrm{He}_n(\phi_i),
  \qquad
  a_{2k+1}=\frac{(-1)^k}{\sqrt{2\pi}\,(2k+1)\,2^k\,k!},\quad a_{2k}=0.
  \label{eq:HalfSpaceHermite}
\end{align}
Hence,
\begin{align}
\Theta(-\phi_{i-1}) \,\Theta(\phi_i)
=\frac14+\sum_{\text{odd }n}\frac{a_n}{2}\big(\mathrm{He}_n(\phi_{i})-\mathrm{He}_n(\phi_{i-1})\big)
  -\sum_{\text{odd }m,n} a_m a_n\,\mathrm{He}_m(\phi_{i-1})\,\mathrm{He}_n(\phi_{i}).
\label{eq:Mi-explicit}
\end{align}

We remark that Eq.~\eqref{eq:Mi-explicit} allows to calculate the mean number of minima $\langle m_N\rangle$. Using $\mathbb{E}[\mathrm{He}_m(X)\mathrm{He}_n(Y)]=\delta_{mn}\,n!\,\rho^{\,n}$ for a standard bivariate normal $(X,Y)$ with correlation $\rho$ (for $\phi_i$, $\rho\equiv\rho_{01}=\mathrm{Corr}(\phi_{i-1},\phi_i)=2^{\,2H-1}-1$), the expectation of  Eq.~\eqref{eq:Mi-explicit} is
\begin{align}
\mathbb{E}\big[\Theta(-\phi_{i-1}) \Theta(\phi_i)\big]
=\frac14-\sum_{k\ge0} a_{2k+1}^2(2k+1)!\,\rho^{\,2k+1}.
\end{align}
Using the series of $\arcsin \rho=\sum_{k\ge0}\frac{(2k)!}{4^k (k!)^2}\frac{\rho^{2k+1}}{2k+1}$ together with
$a_{2k+1}^2(2k+1)! = \frac{1}{2\pi}\frac{(2k)!}{4^k (k!)^2}\frac{1}{2k+1}$, we obtain
\begin{align}
\mathbb{E}\big[\Theta(-\phi_{i-1}) \Theta(\phi_i)\big]
=\frac14-\frac{1}{2\pi}\arcsin\rho\equiv\mu_{\rho}.
\label{eq:mu_rho}
\end{align}
Therefore, for $m_N$ of Eq.~\eqref{m_N_def},
\[
\boxed{\ \langle m_N\rangle=(N-1)\Big(\frac14-\frac{1}{2\pi}\arcsin\rho\Big),\ }
\]
the same result as found previously, see Sec.~\ref{sec:mean}.

Now, we aim to analyze $m_N$ according to the classification of Hermite processes, whose rank is equal to the minimal nonzero order in their algebraic expansion in the Hermite polynomials  (see Ch. 5.2 of Ref.~\cite{PipirasTaqqu2017}). Equation~\eqref{eq:Mi-explicit} contains the term
\[
L_i\equiv\sum_{\text{odd }n}\frac{a_n}{2}\big(\mathrm{He}_n(\phi_i)-\mathrm{He}_n(\phi_{i-1})\big),
\]
whose sum over $i$ is telescopic:
\[
\sum_{i=1}^{N-1}L_i=\sum_{\text{odd }n}\frac{a_n}{2}\big(\mathrm{He}_n(\phi_{N-1})-\mathrm{He}_n(\phi_0)\big),
\]
thus does not contribute to $m_N$ asymptotically as $N\to\infty$. The term of Eq.~\eqref{eq:Mi-explicit} having minimal degree, with non-cancelled in $m_N$ increments $\{\phi_i\}$ for all $i$, is proportional to $\mathrm{He}_1(\phi_{i-1})\,\mathrm{He}_1(\phi_{i})$. This bilinear form can be diagonalized and expressed through the Hermite polynomials $\mathrm{He}_2$. Therefore, the $m_N$ of Eq.~\eqref{m_N_def} has asymptotically ($N\to\infty$)  the Hermite rank $2$. In the following, we will analyze this second-order part.

Let $M_i\equiv\Theta(-\phi_{i-1})\,\Theta(\phi_i)$ and $M_i^\circ\equiv M_i-\langle M_i\rangle=M_i-\mu_{\rho}$.
We decompose
\begin{equation}
M_i^\circ \;=\; Q_i \;+\; R_i 
\end{equation}
where the second–order (Wick) part $Q_i$ is
\begin{equation}
Q_i
= c_1\big(\mathrm{He}_2(\phi_{i-1})+\mathrm{He}_2(\phi_{i})\big)
+ c_2\big(\mathrm{He}_1(\phi_{i-1})\,\mathrm{He}_1(\phi_{i})-\rho\big).
\label{eq:Qi-def}
\end{equation}
Here the cross term is centered ($\mathbb{E}[\phi_{i-1}\phi_i]=\rho$) and the remainder $R_i$ is orthogonal to the whole quadratic Wick span:
\begin{equation}
\langle R_i,\ \mathrm{He}_2(\phi_{i-1})+\mathrm{He}_2(\phi_{i})\rangle=0,
\qquad
\langle R_i,\ \mathrm{He}_1(\phi_{i-1})\mathrm{He}_1(\phi_{i})-\rho\rangle=0.
\end{equation}
Equivalently, $R_i$ contains only Wick components of order $q\neq2$ (the linear part $q=1$ which does not contribute asymptotically to $m_N$ and the higher orders $q\ge3$).

In order to find the coefficients $c_1$ and $c_2$ in Eq.~\eqref{eq:Qi-def}, we introduce the Wick basis vectors
\[
B_1\equiv\mathrm{He}_2(\phi_{i-1})+\mathrm{He}_2(\phi_i)=\phi_{i-1}^2+\phi_i^2-2,\qquad
B_2\equiv\mathrm{He}_1(\phi_{i-1})\mathrm{He}_1(\phi_i)-\rho=\phi_{i-1}\phi_i-\rho,
\]
and the inner product $\langle B_i,B_j\rangle=\mathbb{E}[B_i\,B_j]$. For a standard bivariate normal $(\phi_{i-1},\phi_i)$ with correlation $\rho$, the 
Gram matrix is
\begin{equation}
G=\begin{pmatrix}
\langle B_1,B_1\rangle & \langle B_1,B_2\rangle\\
\langle B_2,B_1\rangle & \langle B_2,B_2\rangle
\end{pmatrix}
=
\begin{pmatrix}
4(1+\rho^2) & 4\rho\\
4\rho & 1+\rho^2
\end{pmatrix}.
\label{eq:Gram-new}
\end{equation}
The projection equations for $Q_i=c_1 B_1+c_2 B_2$ read
\[
\langle M_i^\circ,B_j\rangle=\langle Q_i,B_j\rangle,\qquad j=1,2,
\]
i.e. $G\,c=b$ with $c=(c_1,c_2)^\top$ and $b=(b_1,b_2)^\top$, where
\[
b_1\equiv\langle M_i^\circ,B_1\rangle=\mathbb{E}\!\big[(\mathrm{He}_2(\phi_{i-1})+\mathrm{He}_2(\phi_i))\,\Theta(-\phi_{i-1}) \Theta(\phi_i)\big],\quad
b_2\equiv\langle M_i^\circ,B_2\rangle=\mathbb{E}\!\big[(\phi_{i-1}\phi_i-\rho)\,\Theta(-\phi_{i-1}) \Theta(\phi_i)\big].
\]

We start with the calculation of $b_2$. Consider a bivariate normal pair $(X,Y)$ of correlation $\rho$ with density
\[
f_{\rho}(x,y)=\frac{1}{2\pi\,\sqrt{1-\rho^2}}
\exp\!\Big(-\frac{x^2-2\rho xy+y^2}{2(1-\rho^2)}\Big).
\]
With this,
\begin{align*}
    \partial_\rho\mathbb{E}[\Theta(-X) \Theta(Y)]=\int\Theta(-x) \Theta(y)f_{\rho}(x,y)\partial_\rho \log f_\rho(x,y)\mathrm{d}x\mathrm{d}y.
\end{align*}
Using
\[
\partial_\rho \log f_\rho(x,y)=\frac{xy-\rho}{1-\rho^2},
\]
we get
\begin{align*}
    \partial_\rho\mathbb{E}[\Theta(-X) \Theta(Y)]=\mathbb{E}\bigg[\frac{XY-\rho}{1-\rho^2}\Theta(-X) \Theta(Y)\bigg]=\frac{b_2}{1-\rho^2}.
\end{align*}
From the other hand, $\mathbb{E}[\Theta(-X) \Theta(Y)]=\mu_{\rho}$ and with Eq.~\eqref{eq:mu_rho} we obtain
\begin{equation}
b_2=(1-\rho^2)\,\partial_{\rho}\mu_{\rho}= -\,\frac{1}{2\pi}\sqrt{1-\rho^2}\, .
\label{eq:b2-new}
\end{equation}

To calculate $b_1$, we consider the scaled pair $(X_s,Y_s)=(sX,sY)$ with density
\[
f_{\rho,s}(x,y)=\frac{1}{2\pi\,s^2\sqrt{1-\rho^2}}
\exp\!\Big(-\frac{x^2-2\rho xy+y^2}{2s^2(1-\rho^2)}\Big),
\]
for which
\[
\partial_s\log f_{\rho,s}(x,y)=-\frac{2}{s}+\frac{x^2-2\rho xy+y^2}{s^3(1-\rho^2)}
\ \Longrightarrow\
\partial_s\log f_{\rho,s}\big|_{s=1}=-2+\frac{X^2-2\rho XY+Y^2}{1-\rho^2}
\]
holds. The indicator $\Theta(-X) \Theta(Y)$ is invariant under rescaling with $s$, so that the differentiation at $s=1$ gives 
\[
0=\mathbb{E}\!\big[\partial_s\log f_{\rho,s}(X,Y)\big|_{s=1}\Theta(-X) \Theta(Y)\big]=\mathbb{E}\!\big[\big(\frac{X^2-2\rho XY+Y^2}{1-\rho^2}-2\big)\Theta(-X) \Theta(Y)\big],
\]
leading to $b_1=\mathbb{E}[(X^2-1+Y^2-1)\Theta(-X) \Theta(Y)]=2\rho\mathbb{E}[(XY-\rho)\Theta(-X) \Theta(Y)]=2\rho b_2$.
Therefore
\begin{equation}
b_1=2\rho\,b_2\ =\ -\,\frac{\rho}{\pi}\,\sqrt{1-\rho^2}\ .
\label{eq:b1-new}
\end{equation}

Now we have all coefficients of the system $G\,c=b$ on $c_1$ and $c_2$ that explicitly reads
\begin{equation}
  \begin{cases}
4(1+\rho^2)\,c_1 + 4\rho\,c_2 = -\frac{\rho}{\pi}\sqrt{1-\rho^2},\\
4\rho\,c_1 + (1+\rho^2)\,c_2 = -\frac{1}{2\pi}\sqrt{1-\rho^2}.
\end{cases}  
\end{equation}
Its solution is 
\begin{equation}
c_1=\frac{\rho}{4\pi\,\sqrt{1-\rho^2}},\quad c_2= -\frac{1}{2\pi\,\sqrt{1-\rho^2}}.
\label{eq:c1c2}
\end{equation}
Summarizing, the second–order (Wick) part $Q_i$ of the centered local minima indicator $M_i^{\circ}$ is
\begin{equation}
\boxed{
Q_i
= \frac{\rho}{4\pi\sqrt{1-\rho^2}}\big[\mathrm{He}_2(\phi_{i-1})+\mathrm{He}_2(\phi_{i})\big]
-\frac{1}{2\pi\sqrt{1-\rho^2}}\big[\mathrm{He}_1(\phi_{i-1})\,\mathrm{He}_1(\phi_{i})-\rho\big].}
\label{eq:Qi-final}
\end{equation}

Define the sum/difference modes
\[
V_i=\frac{\phi_i+\phi_{i-1}}{\sqrt{2(1+\rho)}},
\qquad
U_i=\frac{\phi_i-\phi_{i-1}}{\sqrt{2(1-\rho)}},
\]
so that $U_i,V_i$ are standard and independent (at fixed $i$). Then
\[
\mathrm{He}_2(\phi_{i-1})+\mathrm{He}_2(\phi_i)=(1+\rho)\,\mathrm{He}_2(V_i)+(1-\rho)\,\mathrm{He}_2(U_i),
\]
\[
\phi_{i-1}\phi_i-\rho=\frac{1+\rho}{2}\,\mathrm{He}_2(V_i)-\frac{1-\rho}{2}\,\mathrm{He}_2(U_i),
\]
the second–order (Wick) part $Q_i$ becomes $Q_i=g_V\,\mathrm{He}_2(V_i)+g_U\,\mathrm{He}_2(U_i)$, with
\begin{align}\label{eq:gVgU}
 g_V=(1+\rho)\Big(c_1+\frac{c_2}{2}\Big)=-\,\frac{\sqrt{1-\rho^2}}{4\pi},\quad
g_U=(1-\rho)\Big(c_1-\frac{c_2}{2}\Big)=\,\frac{\sqrt{1-\rho^2}}{4\pi}.   
\end{align}

Summing $Q_i$, we get
\begin{equation}
\boxed{\
m_N-\langle m_N\rangle
= -\frac{\sqrt{1-\rho^2}}{4\pi}\sum_{i=1}^{N-1}\mathrm{He}_2(V_i)
+ \frac{\sqrt{1-\rho^2}}{4\pi}\sum_{i=1}^{N-1}\mathrm{He}_2(U_i)
\ +\sum_{i=1}^{N-1}R_i.
}
\label{eq:fluct-decomp}
\end{equation}
The sum of the remainder $R_i$ contains (i) boundary increments that do not contribute asymptotically to ${m_N-\langle m_N\rangle}$ as $N\to\infty$ at any $H$ and 
(ii) the fluctuations of higher than quadratic order that are irrelevant for $H>3/4$ (Theorem~5.3.3 of Ref.~\cite{PipirasTaqqu2017}).

We are going now to concentrate on the regime (ii), in which the high-$i$ terms linear in $\mathrm{He}_2(\dots)$ dominate the sum in Eq.~\eqref{eq:fluct-decomp}. For fBm $\rho_{0k}=\mathbb{E}[\phi_0\phi_k]\sim C_H\,k^{2H-2}$ with $C_H=H(2H-1)$. A direct computation gives, for $k\ge1$,
\begin{align}
\mathrm{Cov}(V_1,V_{1+k})
&=\frac{\mathbb{E}\big[(\phi_1+\phi_{0})(\phi_{k+1}+\phi_{k})\big]}{2(1+\rho)}
=\frac{2\rho_{0k}+\rho_{0,k-1}+\rho_{0,k+1}}{2(1+\rho)},\label{eq:covV-anch1}\\[2pt]
\mathrm{Cov}(U_1,U_{1+k})
&=\frac{\mathbb{E}\big[(\phi_1-\phi_{0})(\phi_{k+1}-\phi_{k})\big]}{2(1-\rho)}
=\frac{2\rho_{0k}-\rho_{0,k-1}-\rho_{0,k+1}}{2(1-\rho)}.\label{eq:covU-anch1}
\end{align}
Using the expansion $\rho_{0,k\pm1}=C_H (k\pm1)^{2H-2}
=C_H\big[k^{2H-2}\pm(2H-2)k^{2H-3}+\tfrac{(2H-2)(2H-3)}{2}k^{2H-4}+o(k^{2H-4})\big]$,
we obtain, as $k\to\infty$,
\begin{align}\label{eq:covV1-asym}
\mathrm{Cov}(V_1,V_{k+1})
=\frac{2C_H}{1+\rho}\,k^{2H-2}+O(k^{2H-4}),
\qquad
\mathrm{Cov}(U_1,U_{k+1})
=-\frac{C_H(2H-2)(2H-3)}{2(1-\rho)}\,k^{2H-4}+o(k^{2H-4}).
\end{align}
Thus the $V$-mode inherits the long memory $\propto k^{2H-2}$, whereas the $U$-mode is two derivatives shorter, at $\propto k^{2H-4}$. In particular, using that $\mathrm{Cov}(\mathrm{He}_2(V_1),\mathrm{He}_2(V_{k+1}))=2\,\mathrm{Cov}(V_1,V_{k+1})^2$, the fluctuations of $m_N-\langle m_N\rangle$ are dominated by the $\{V_i\}$ leading for $H>\tfrac34$ to the variance $\mathrm{Var}(m_N)$ of order $N^{4H-2}$. By the Dobrushin–Major–Taqqu theorem for long-range dependent Gaussian functionals of Hermite rank $2$ \cite{PipirasTaqqu2017,DobrushinMajor1979,Taqqu1979}, we conclude that, for $H>\tfrac34$,
\begin{align}\label{eq:rosenblatt}
\frac{\big(m_N-\langle m_N\rangle\big)}{\sqrt{\mathrm{Var}(m_N)}} \Longrightarrow\,-\mathcal{R},
\end{align}
where $\mathcal{R}$ is the canonical Rosenblatt random variable (of unit variance)
and the negative sign comes from the negative prefactor of the first rhs term of Eq.~\eqref{eq:fluct-decomp} which dominates the process. The result \eqref{eq:rosenblatt} is in line with the Slud theorem \cite{Slud1994} for the number of zero crossings of a unit interval by a continuous process, expressed as the second-order multiple Wiener-Itô integral. The bridge to our result passes through the variance $\mathrm{Var}(m_N)$ and the proper sign of the Rosenblatt process, keeping in mind its scale-invariance. 

\begin{figure}
    \captionsetup{justification=raggedright} 
    \centering
    \begin{subfigure}[t]{0.32\textwidth}
        \begin{overpic}[width=\linewidth]{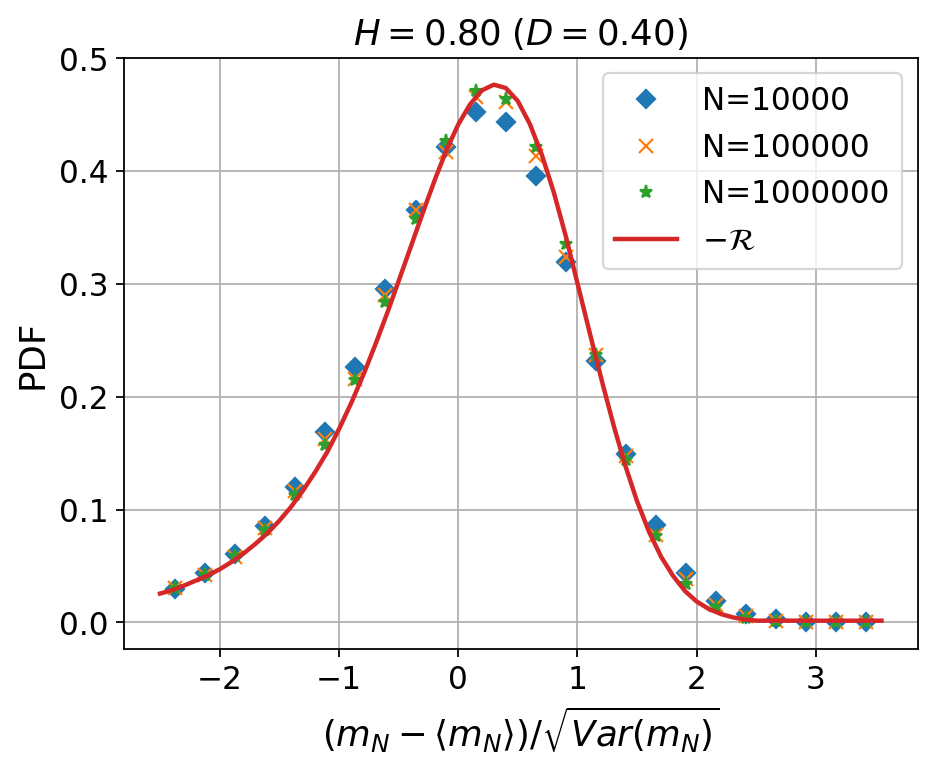}
            \put(0,130){\textbf{(a)}}
        \end{overpic}
    \end{subfigure}
    \hfill
    \begin{subfigure}[t]{0.32\textwidth}
        \begin{overpic}[width=\linewidth]{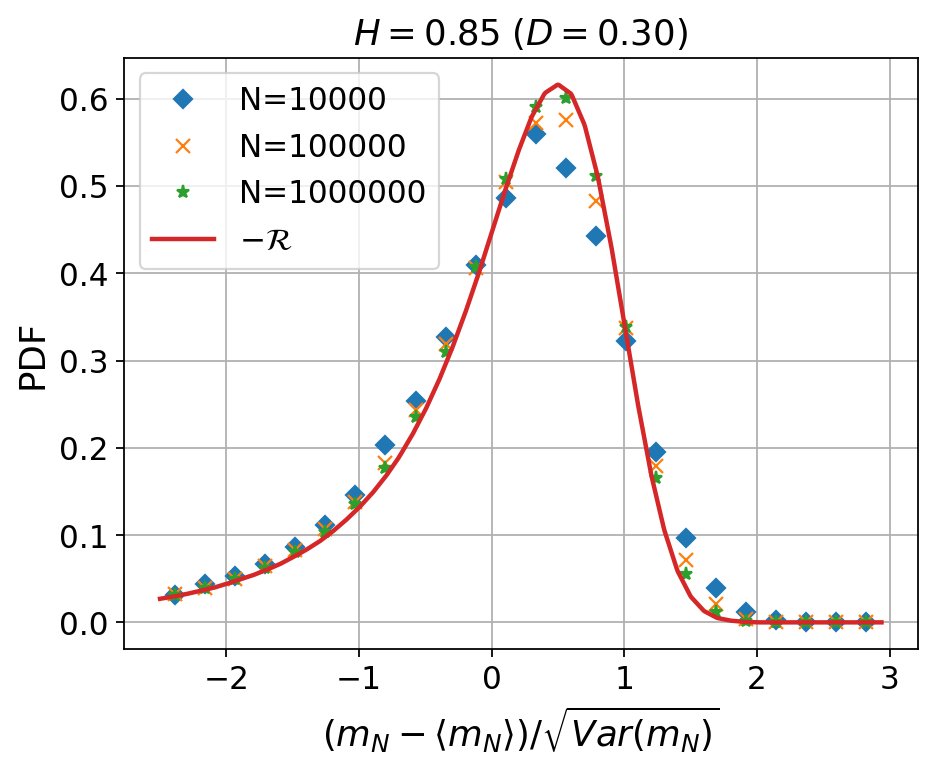}
             \put(0,130){\textbf{(b)}}
        \end{overpic}
    \end{subfigure}
    \hfill
    \begin{subfigure}[t]{0.32\textwidth}
        \begin{overpic}[width=\linewidth]{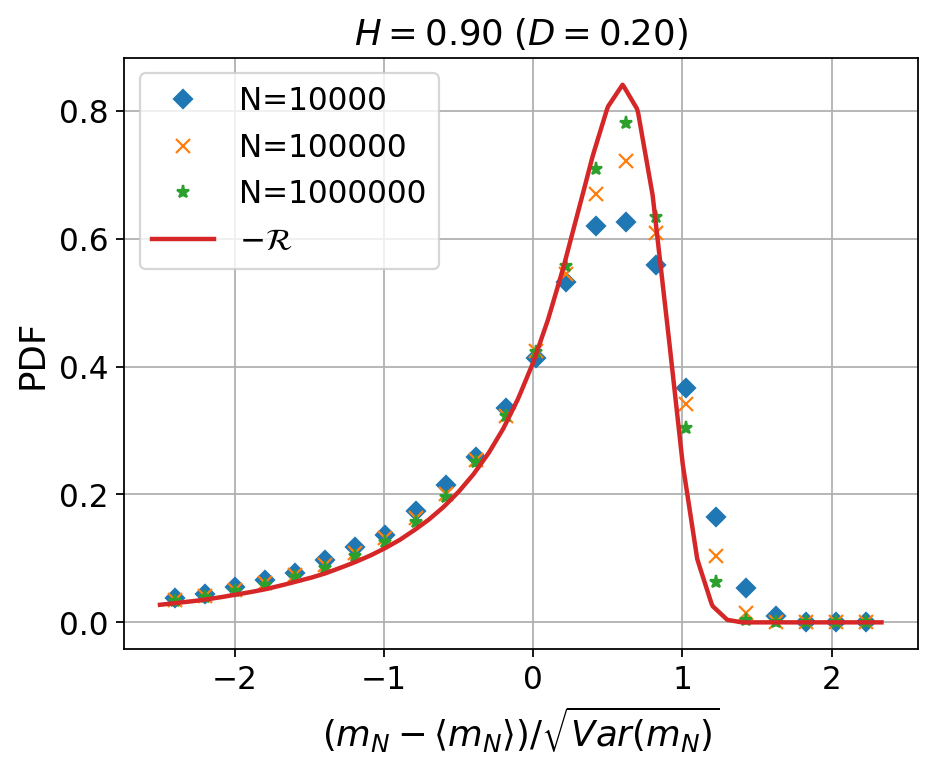}
             \put(0,130){\textbf{(c)}}
        \end{overpic}
    \end{subfigure}
\caption{The pdf of centered and normalized number of minima $m_N$ obtained from fBm simulations for different values of $N$ compared with the Rosenblatt pdf (from Ref.~\cite{VeilletteTaqqu2013}). Different subfigures are for different values of $H$ related to the Rosenblatt index as  $D=2-2H$. For higher values of $H$ the approach to the Rosenblatt asymptotics is very sluggish.}
\label{fig:rosenblatt}
\end{figure}

A closed analytic formula for the Rosenblatt distribution is unknown, its computation is a hard task and an active topic in mathematical literature \cite{VeilletteTaqqu2013,leonenko2025numerical}. Here we use the numerically obtained Rosenblatt pdf based on the tabulated values in the supplemental material of Ref.~\cite{VeilletteTaqqu2013}. In Fig.~\ref{fig:rosenblatt} we show the convergence of the centered and normalized number of minima obtained from fBm simulations for different values of $N$ to the Rosenblatt pdf. We note that the higher is the value of the Hurst exponent, the slower is the convergence.

The asymptotic variance $\mathrm{Var}(m_N)$ for $H>3/4$ is calculated in the previous section (see Eq.~\eqref{variance_fbm_fast} which corrects the result of Ref.\cite{SinnKeller2011}). Here we obtain the same result from Eq.~\eqref{eq:fluct-decomp}. Let us note
\[
S_N^{(V)}=\sum_{i=1}^{N-1}\mathrm{He}_2(V_i),\qquad
S_N^{(U)}=\sum_{i=1}^{N-1}\mathrm{He}_2(U_i).
\]
Then
\[
\mathrm{Var}(m_N)=\frac{1-\rho^2}{16\pi^2}\left[\mathrm{Var}(S_N^{(V)})+\mathrm{Var}(S_N^{(U)})
-2\mathrm{Cov}(S_N^{(V)},S_N^{(U)})\right]+O(1).
\]
The cross term and the $U$-term are $O(N)$ and negligible for $H>\tfrac34$.
For the dominant $V$-part,
\[
\mathrm{Var}(S_N^{(V)})=(N-1)\,\mathrm{Var}(\mathrm{He}_2(V_1))
+2\sum_{k=1}^{N-2}(N-1-k)\,\mathrm{Cov}(\mathrm{He}_2(V_1),\mathrm{He}_2(V_{k+1})).
\]
Since $\mathrm{Var}(\mathrm{He}_2(V_1))=2$ and
$\mathrm{Cov}(\mathrm{He}_2(V_1),\mathrm{He}_2(V_{1+k}))=2\,\mathrm{Cov}(V_1,V_{1+k})^2$,
using Eq.~\eqref{eq:covV1-asym} we get for $H>\tfrac34$,
\[
\mathrm{Var}(S_N^{(V)})\sim 4\left(\frac{2C_H}{1+\rho}\right)^2\,\sum_{k=1}^{N} (N-k)\,k^{4H-4}\sim \frac{16C_H^2}{(1+\rho)^2(4H-3)(4H-2)}\,N^{4H-2}.
\]
Therefore,
\[
\mathrm{Var}(m_N)\ \sim\ \dfrac{1-\rho^2}{16\pi^2}\,\mathrm{Var}(S_N^{(V)})
\sim\ \Bigg[\ \frac{1}{\pi^2}\,\frac{1-\rho}{1+\rho}\,
\frac{C_H^{\,2}}{(4H-3)(4H-2)}\ \Bigg]\ N^{4H-2}.
\]
Substituting $C_H=H(2H-1)$ and $\rho=2^{\,2H-1}-1$ yields
\begin{equation}
\boxed{\
\mathrm{Var}(m_N)\ \sim \frac{4^{1-H}-1}{2\pi^2}\frac{H^2(2H-1)}{4H-3} \ N^{4H-2},
}
\end{equation}
which is Eq.~\eqref{variance_fbm_fast} of Sec. I and Eq.~(6) of the main text.

Coming back to the regime of shorter range correlated random variables, $H\le 3/4$, the Breuer–Major theorem \cite{PipirasTaqqu2017,BreuerMajor1983} states that the fluctuations of $m_N$ of Eq.~\eqref{eq:fluct-decomp} are Gaussian and the centered and normalized $m_N$ converges to the Brownian motion $\mathcal{B}$ \cite{PipirasTaqqu2017,GiraitisSurgailis1985,AzmoodehPeccatiPoly2016},
\begin{equation}
\frac{m_N-\langle m_N\rangle}{\sqrt{\mathrm{Var}(m_N)}} \Longrightarrow \mathcal{B}.
\label{eq:brownian}
\end{equation}
We note that in the marginally non-summable case $H=3/4$ a CLT still holds (but with unusual normalization $\sqrt{N\log N}$), with logarithmically slow convergence to the normal distribution \cite{GiraitisSurgailis1985}.

Importantly, Eqs.~\eqref{eq:rosenblatt} and \eqref{eq:brownian} represent the convergence in the process, which provide us the full statistics of $m_N$ at any number of times. Here we exemplify it on the case of two times by considering the covariance  $\mathbb{E}[Z_K(t)Z_K(s)]$ of the process
\begin{align}
    Z_K(t)=\frac{m_{[Kt]}-\langle m_{[Kt]}\rangle}{\sqrt{\mathrm{Var}(m_K)}},
\end{align}
see Fig. 3 in the main text, and compare it for $H>3/4$ with the covariance of the Rosenblatt process $\mathcal{R}$ \cite{maejima2007wiener}:
\begin{align}
    \mathbb{E}[\mathcal{R}(t)\mathcal{R}(s)]=\frac{1}{2}\left(t^{2(2H-1)}+s^{2(2H-1)}-|t-s|^{2(2H-1)}\right)
\end{align}
and for $H\leq3/4$ with this of the Brownian motion $\mathbb{E}[\mathcal{B}(t)\mathcal{B}(s)]=\min(t,s)$ \cite{PipirasTaqqu2017}. We note that for $H=3/4$ the convergence of $\mathbb{E}[Z_K(t)Z_K(s)]$ to the Brownian result is logarithmically slow, $\mathbb{E}[Z_K(t)Z_K(s)]\sim\min(t,s)+\left(t\ln t+s\ln s-|t-s|\ln|t-s|\right)/(2\ln K) + \mathcal{O}((\ln K)^{-2})$.